\documentclass[sigconf, screen, nonacm]{acmart}

\makeatletter
\def\@ACM@checkaffil{
    \if@ACM@instpresent\else
    \ClassWarningNoLine{\@classname}{No institution present for an affiliation}
    \fi
    \if@ACM@citypresent\else
    \ClassWarningNoLine{\@classname}{No city present for an affiliation}
    \fi
    \if@ACM@countrypresent\else
        \ClassWarningNoLine{\@classname}{No country present for an affiliation}
    \fi
}
\makeatother

\AtBeginDocument{
  \providecommand\BibTeX{{
    \normalfont B\kern-0.5em{\scshape i\kern-0.25em b}\kern-0.8em\TeX}}}

\newcommand{\diff}[1]{\textcolor{red}{#1}}

\newtheorem{theorem}{Theorem}

\newenvironment{proof1} {\begin{proof}[Justification]} {\end{proof}}

\usepackage{paralist}
\usepackage{afterpage}
\usepackage{multirow}

\usepackage[ruled,vlined]{algorithm2e}

\usepackage{listings}
\usepackage[noend]{algpseudocode}
\usepackage{url}
\usepackage{multirow}
\usepackage[font={small,it}]{caption}
\usepackage{array}
\usepackage{rotating}   
\usepackage{makecell}
\usepackage{colortbl}
\usepackage{enumerate}
\usepackage{booktabs}
\usepackage{subfig}
\usepackage{commath}
\usepackage{amsthm}
\usepackage{bm}
\usepackage{framed}
\usepackage[bottom]{footmisc}

\usepackage[misc]{ifsym}
\usepackage[utf8x]{inputenc}

\algrenewcommand\algorithmiccomment[1]{\hfill \textcolor{gray}{$\triangleright$ \textit{#1}}}

\begin{document}

\title{CBDC-AquaSphere: Interoperable Central Bank Digital Currency Built on Trusted Computing and Blockchain}

\settopmatter{authorsperrow=1} 
\newcommand{\tsc}[1]{\textsuperscript{#1}} 

 \author{Ivan Homoliak,$^\dagger$ Martin Pere\v{s}\'{i}ni,$^\dagger$ Patrik Holop,$^\dagger$ Jakub Handzu\v{s},$^\dagger$  Fran Casino\tsc{$\ast$}}

 \affiliation{
 	\vspace{0.4cm}
   \institution{$^\dagger$Brno University of Technology, Faculty of Information Technology, Czech Republic}
   \institution{$^\ast$Universitat Rovira i Virgili, Tarragona, Spain}
 }

\begin{abstract}
The adoption of decentralized, tamper-proof ledger systems is paving the way for new applications and opportunities in different contexts.
While most research aims to improve their scalability, privacy, and governance issues, interoperability has received less attention. 
Executing transactions across various blockchains is notably instrumental in unlocking the potential of novel applications, particularly in the financial sector, where their potential would otherwise be significantly diminished.
Therefore, interoperable ledgers are crucial to ensure the expansion and further adoption of such a technology in various contexts. 

In this paper, we present a protocol that uses a combination of trusted execution environment (TEE) and blockchains to enable interoperability  over independent semi-centralized CBDC ledgers, guaranteeing the atomicity of inter-bank transfers. 
Our interoperability protocol uses a custom adaptation of atomic swap protocol and is executed by any pair of CBDC instances to realize a one-way transfer.
It ensures features such as atomicity, verifiability, correctness, censorship resistance, and privacy while offering high scalability in terms of the number of CBDC instances.
Our approach enables to possible deployment scenarios that can be combined: 
(1) CBDC instances represent central banks of multiple countries, and 
(2) CBDC instances represent the set of retail banks and a paramount central bank of a single country.
We provide a detailed description of our protocol as well as an extensive analysis of its benefits, features, and security.

In this WIP paper, we made a proof-of-concept implementation and made a partial evaluation, while the more extensive evaluation will be made in our future work.

\end{abstract}


\maketitle

\section{Introduction}
\label{sec:intro}

Blockchain technology is becoming the backbone of a myriad of applications since it provides features such as decentralization, immutability, availability, and transparency. 
More recently, along with the increasing adoption and maturity of such a technology~\cite{alshamsi2022systematic}, central banks all over the world are accelerating the process of Central Bank Digital Currency (CBDC) development~\cite{zhang2021blockchain}. CBDC has received increasing attention in the past few years. More than 85\% of central banks are actively researching the potential for CBDCs, and according to BIS survey~\cite{boar2021ready} conducted in 2021 central banks covering 20\% of the world's population are likely to launch retail CBDCs before 2025. 
Some of the reasons behind this new paradigm are the digitization of the economy, the level of development of the financial sector, and a strong decline in the use of cash~\cite{nanez2020reasons}.

Despite the generalized will to improve the worldwide financial system by utilizing blockchain technology~\cite{swift_link} in centralized environments, there is still a road ahead for the realization of fast secure blockchain payment systems. 
Nevertheless, some features are essential to enable financial solutions to reach an operational level, making interoperability a crucial requirement in this context. 
Note that a few cross-chain solutions and protocols~\cite{zhang2021blockchain,lan2021trustcross,bellavista2021interoperable} that leverage the necessary level of interoperability for execution of inter-bank financial transactions have been proposed in the literature. 
In this regard, technologies such as Trusted Execution Environments (TEE) in a potential combination with blockchains can efficiently enforce the required security and privacy levels of centralized environments of banks, and thus provide a high level of trustworthiness for the end users.

\paragraph{\textbf{Motivation}}
CBDC legislation and adoption goes in hand with privacy and security concerns. The centralized nature of banks implies that transactions are recorded in private ledgers managed by banks, contrary to the very nature of public decentralized cryptocurrencies. 
While this may prevent some potential malicious scenarios, users are forced into trusting a single authority and its corresponding regulations. 
Aiming at increasing decentralization and trust, several authors have proposed the use of TEE to leverage verifiable protocols enabling interoperability of multiple centralized isolated environments \cite{wang2023exploring,lacoste2023trusted}. 

While achieving blockchain interoperability is challenging regardless of its flavor (i.e., between centralized, decentralized or hybrid blockchain structures) additional features such as scalability, confidentiality, and censorship resistance are necessary to guarantee for practical scenarios.
Digital Euro Association released on October 2022 the CBDC manifesto~\cite{cbdc-manifesto}, in which they highlight important features of CBDC, such as strong value proposition for the end users, the highest degree of privacy, and interoperability. 
Our work is inline with this manifesto and adheres to its features while it provides even additional features that are interesting for the users and the whole ecosystem assuming that the CBDC-equipped bank might potentially  be an untrusted entity.

\paragraph{\textbf{Contributions}}
In this paper, we present a practical blockchain interoperability protocol that integrates such features. On top of the above mentioned features, to the best of our knowledge, our work represents the first TEE-based interoperable CBDC approach that provides the proof-of-censorship. 
Our main contributions are summarized as follows: 

\begin{compactenum}

    \item We specify requirements for an instance of CBDC that is controlled by a single bank\footnote{As we will see it can be a central bank or even a retail bank, depending on the deployment scenario described in \autoref{sec:identity-management-of-CBDC}. Therefore, depending on the scenario, we well use the term CBDC instance even for a retail bank.} and forms an isolated environment.
    These requirements include high processing performance, transparent token issuance, correctness of intra-bank transfers, immutability of historical data, non-equivocation, privacy, and the indisputable proofs of censorship.
    
    \item We investigate state-of-the-art approaches applicable for a CBDC instance assuming our requirements and identify the most convenient one, Aquareum~\cite{homoliak2020aquareum}, that we further base on.
	The Aquareum-based CBDC ledger ensures immutability, non-equivocation, privacy, and the indisputable proofs of transaction censorship by utilizing a permissionless blockchain (e.g., Ethereum).
	Next, by using TEE  (e.g., Intel SGX), it ensures the correctness of any transaction execution.
	
	\item Our main contribution resides in a design and implementation of a protocol resolving interoperability over multiple instances of semi-centralized CBDC, which guarantees the atomicity of inter-bank transfers. 
	
	\item We provide a security analysis, 
	to prove the the properties of our approach. 
\end{compactenum}

\paragraph{\textbf{Organization}}
The remainder of the article is organized as follows. 
In \autoref{sec:background}, we provide a background on blockchain, atomic swap, and trusted computing. 
We define the problem in \autoref{sec:problem}, where we describe the attacker model and required features of a single CBDC instance as well as the environment of multiple interoperable CBDC instances.
\autoref{sec:design} provides a description of the proposed interoperability protocol and its deployment scenarios. 
Next, the implementation details of the designed protocol and its partial evaluation are described in \autoref{sec:eval}. 
We make a security analysis of our approach in \autoref{sec:secanalysis}.
\autoref{sec:related} reviews the state of the art of CBDC approaches and TEE-based blockchain solutions. 
\autoref{sec:discussion} discusses the benefits and limitations of our approach. 
We conclude the paper in \autoref{sec:conclusion} with some final remarks.

 \section{Background}
 \label{sec:background}
 This section provides the reader with the essential context needed to understand the topics that will be discussed in this article.

\subsection{Blockchain}\label{sec:background-blockchain}
Blockchain is a tamper-resistant data structure, in which data records (i.e., \textit{blocks}) are linked using a cryptographic hash function, and each new block has to be agreed upon by participants (a.k.a., \textit{miners}) running a consensus protocol (i.e., \textit{consensus nodes}).
Each block may contain data records representing orders that transfer tokens, application codes written in a platform-supported language, and the execution orders of such application codes. 
These application codes are referred to as \textit{smart contracts}, and they encode arbitrary processing logic written in a supported language of a smart contract platform.
Interactions between clients and the smart contract platform are based on messages called \textit{transactions}.

\subsection{Trusted Execution Environment}
Trusted Execution Environment (TEE) is a hardware-based component that enables secure (remote) execution~\cite{subramanyan2017formal} of a pre-defined code (i.e., enclave) in an isolated environment. 
TEE uses cryptographic primitives and hardware-embedded secrets that protect data confidentiality and the integrity of computations.
In particular, the adversary model of TEE involves the operating system (OS) that may compromise user-space applications but not TEE-protected applications.
An enclave process cannot execute system calls but can read and write memory outside the enclave. 
Thus isolated execution in TEE may be viewed as an ideal model in which a process is guaranteed to be executed correctly with ideal confidentiality, while it might run on a potentially malicious OS.

\paragraph{\textbf{Intel SGX}}
While there exist multiple instances of TEE, in the context of this work we will focus on Intel SGX (Software Guard Extensions)~\cite{anati2013innovative,mckeen2013innovative,hoekstra2013using}.
Intel SGX allows a local process or a remote system to securely communicate with the enclave as well as execute verification of the integrity of the enclave's code. 
When an enclave is created, the CPU outputs a report of its initial state, also referred to as a \textit{measurement}, which is signed by the private key of TEE and encrypted by a public key of Intel Attestation Service (IAS). 
The hardware-protected signature serves as the proof that the measured code is running in an SGX-protected enclave, while the encryption by IAS public key ensures that the SGX-equipped CPU is genuine and was manufactured by Intel.
This proof is also known as a \textit{quote} or \textit{attestation}, and it can be verified by a local process or by a remote system. 
The enclave-provided public key can be used by a verifier to establish a secure remote channel with the enclave or to verify the signature during the attestation.

\subsection{CBDC}
CBDC is often defined as a digital liability backed and issued by a central bank that is widely available to the general public. 
CBDC encompasses many potential benefits such as efficiency and resiliency, flexible monetary policies, and enables enhanced control of tax evasion and money laundering~\cite{kiff2020survey}. 
However, regulations, privacy and identity management issues, as well as design vulnerabilities are potential risks that are shared with cryptocurrencies. 
Many blockchain-based CBDC projects rely on using some sort of stable coins adapting permissioned blockchains due to their scalability and the capability to establish specific privacy policies, as compared to public blockchains~\cite{sethaput2021blockchain,zhang2021blockchain}. 
Therefore, the level of decentralization and coin volatility are two main differences between blockchain-based CBDCs and common cryptocurrencies.
These CBDCs are often based on permissioned blockchain projects such as Corda~\cite{brown2016corda}, variants of Hyperledger~\cite{hyperledger-github}, and Quorum~\cite{espel2017proposal}.

CDBC solutions are often designed as multi-layer projects~\cite{2022cbdctypes}. 
Wholesale CBDC targets communication of financial institutions and inter-bank settlements. 
Retail CBDC includes accessibility to the general public or their customers. 

\subsection{Atomic Swap}\label{sec:atomicswap}
A basic atomic swap assumes two parties $\mathbb{A}$ and $\mathbb{B}$ owning crypto-tokens in two different blockchains.
$\mathbb{A}$ and $\mathbb{B}$ wish to execute cross-chain exchange atomically and thus achieve a \textit{fairness} property, i.e., either both of the parties receive the agreed amount of crypto-tokens or neither of them.
First, this process involves an agreement on the amount and exchange rate, and second, the execution of the exchange itself.

In a centralized scenario~\cite{micali2003simple}, the approach is to utilize a trusted third party for the execution of the exchange.
In contrast to the centralized scenario, blockchains allow us to execute such an exchange without a requirement of the trusted party.
The atomic swap protocol~\cite{atomic-swap} enables conditional redemption of the funds in the first blockchain to $\mathbb{B}$ upon revealing of the hash pre-image (i.e., secret) that redeems the funds on the second blockchain to $\mathbb{A}$.
The atomic swap protocol is based on two Hashed Time-Lock Contracts (HTLC) that are deployed by both parties in both blockchains.

Although HTLCs can be implemented by Turing-incomplete smart contracts with support for hash-locks and time-locks, for clarity, we provide a description assuming Turing-complete smart contracts, requiring four transactions:
\begin{enumerate}
	\item $\mathbb{A}$ chooses a random string $x$ (i.e., a secret) and computes its hash $h(x)$.
	Using $h(x)$, $\mathbb{A}$ deploys $HTLC_\mathbb{A}$ on the first blockchain and sends the agreed amount to it, which later enables anybody to do a conditional transfer of that amount to $\mathbb{B}$ upon calling a particular method of $HTLC_\mathbb{A}$ with $x = h(x)$ as an argument (i.e., hash-lock). 
	Moreover, $\mathbb{A}$ defines a time-lock, which, when expired, allows $\mathbb{A}$ to recover funds into her address by calling a dedicated method: this is to prevent aborting of the protocol by another party.
	
	\item When $\mathbb{B}$ notices that $HTLC_\mathbb{A}$ has been already deployed, she deploys $HTLC_\mathbb{B}$ on the second blockchain and sends the agreed amount there, enabling a conditional transfer of that amount to $\mathbb{A}$ upon revealing the correct pre-image of $h(x)$ ($h(x)$ is visible from already deployed $HTLC_\mathbb{A}$).
	$\mathbb{B}$ also defines a time-lock in $HTLC_\mathbb{B}$ to handle abortion by $\mathbb{A}$.
	
	\item Once $\mathbb{A}$ notices deployed $HTLC_\mathbb{B}$, she calls a method of $HTLC_\mathbb{B}$ with revealed $x$, and in turn, she obtains the funds on the second blockchain.
	
	\item Once $\mathbb{B}$ notices that $x$ was revealed by $\mathbb{A}$ on the second blockchain, she calls a method of $HTLC_\mathbb{A}$ with $x$ as an argument, and in turn, she obtains the funds on the first blockchain.
\end{enumerate}
If any of the parties aborts, the counter-party waits until the time-lock expires and redeems the funds.

\subsection{Merkle Tree}\label{sec:MT-background}
A Merkle tree~\cite{merkle1989certified} is a  data structure based on the binary tree in which each leaf node contains a hash of a single data block, while each non-leaf node contains a hash of its concatenated children.
Hence, the root node provides a tamper-evident integrity snapshot of the tree contents.
A Merkle tree enables efficient membership verification (with logarithmic time/space complexity) using the Merkle \textit{proof}.
To enable a membership verification of element $x_i$ in the list $X$, the Merkle tree supports the following operations:
\begin{compactdesc}
	\item{$\mathbf{MkRoot(X) \rightarrow  Root}$:} an aggregation of all elements of the list $X$ by a Merkle tree, providing a single value $Root$. 
	
	\item{$\mathbf{MkProof(x_i, X) \rightarrow  \pi^{mk}}$:} a Merkle proof generation for the $i$th element $x_i$ present in the list of all elements $X$. 
	
	\item{$\mathbf{\pi^{mk}.Verify(x_i, Root) \rightarrow  \{T, F\}}$:}  verification of the Merkle proof $\pi^{mk}$, witnessing that $x_i$ is included in the list $X$ that is aggregated by the Merkle tree with the root hash  $Root$.
\end{compactdesc}

\subsection{History Tree}\label{sec:background-historyT} 
A Merkle tree has been primarily used for proving membership.
However, Crosby and Wallach~\cite{crosby2009efficient} extended its application for an append-only tamper-evident log, named a \textit{history tree}.
In history tree, leaf nodes are added in an append-only fashion while it enables to produce incremental proofs witnessing that arbitrary two versions of the tree are consistent. 
The history tree brings a versioned computation of hashes over the Merkle tree, enabling to prove that different versions (i.e., commitments) of a log, with distinct root hashes, make consistent claims about the past. 
The history tree $L$ supports the following operations:
\begin{compactdesc}
	\item[$\mathbf{L.add(x) \rightarrow C_j}$:] appending of the record $x$ to $L$, returning a new commitment $C_j$ that represents the most recent value of the root hash of the history tree.
	
	\item[$\mathbf{L.IncProof(C_i, C_j) \rightarrow \pi^{inc}}$:] an incremental proof generation between two commitments $C_i$ and $C_j$, where $i \leq j$.
	
	\item[$\mathbf{L.MemProof(i, C_j) \rightarrow \pi^{mem} }$:] a membership proof generation for $x_i$ from the commitment $C_j$, where $i \leq j$. 
	
	\item[$\mathbf{\pi^{inc}.Verify(C_i, C_j) \rightarrow \{T, F\}}$:]  verification of the incremental proof $\pi^{inc}$, witnessing that the commitment $C_j$ contains the same history of records $x_k, k \in \{0,\ldots,i\}$ as the commitment $C_i$, where $i \leq j$.  
	
	\item[$\mathbf{\pi^{mem}.Verify(i, x_i, C_j) \rightarrow \{T, F\}}$:]  verification of the membership proof $\pi^{mem}$, witnessing that $x_i$ is the $i$th record in the $j$th version of $L$, fixed by the commitment $C_j$,  $i \leq j$. 
	
	\item[$\mathbf{\pi^{inc}.ReduceRoot() \rightarrow C_j}$:] a reduction of the commitment $C_j$ from the incremental proof $\pi^{inc}$ that was generated by $L.Inc\-Proof(C_i, C_j)$.
\end{compactdesc}

\subsection{Aquareum}
Aquareum~\cite{homoliak2020aquareum} is a centralized ledger that is based on a combination of a trusted execution environment (TEE) with a public blockchain platform (see our other submission with ID $\#83$). 
It provides a publicly verifiable non-equivocating censor\-ship-evident private
ledger. 
Aquareum is integrated with a Turing-complete virtual machine (instantiated by eEVM~\cite{eEVM-Microsoft}),
allowing arbitrary transaction processing logic, such as transfers or
client-specified smart contracts. 
In other words, Aquareum provides most of the  blockchain features while being lightweight and cheap in contrast to them. 
Nevertheless, Aquareum does not provide extremely high availability (such as blockchains) due to its centralized nature, which is, however, common and acceptable for the environment of CBDC.

\begin{figure}[t]
	\includegraphics[width=0.85\columnwidth]{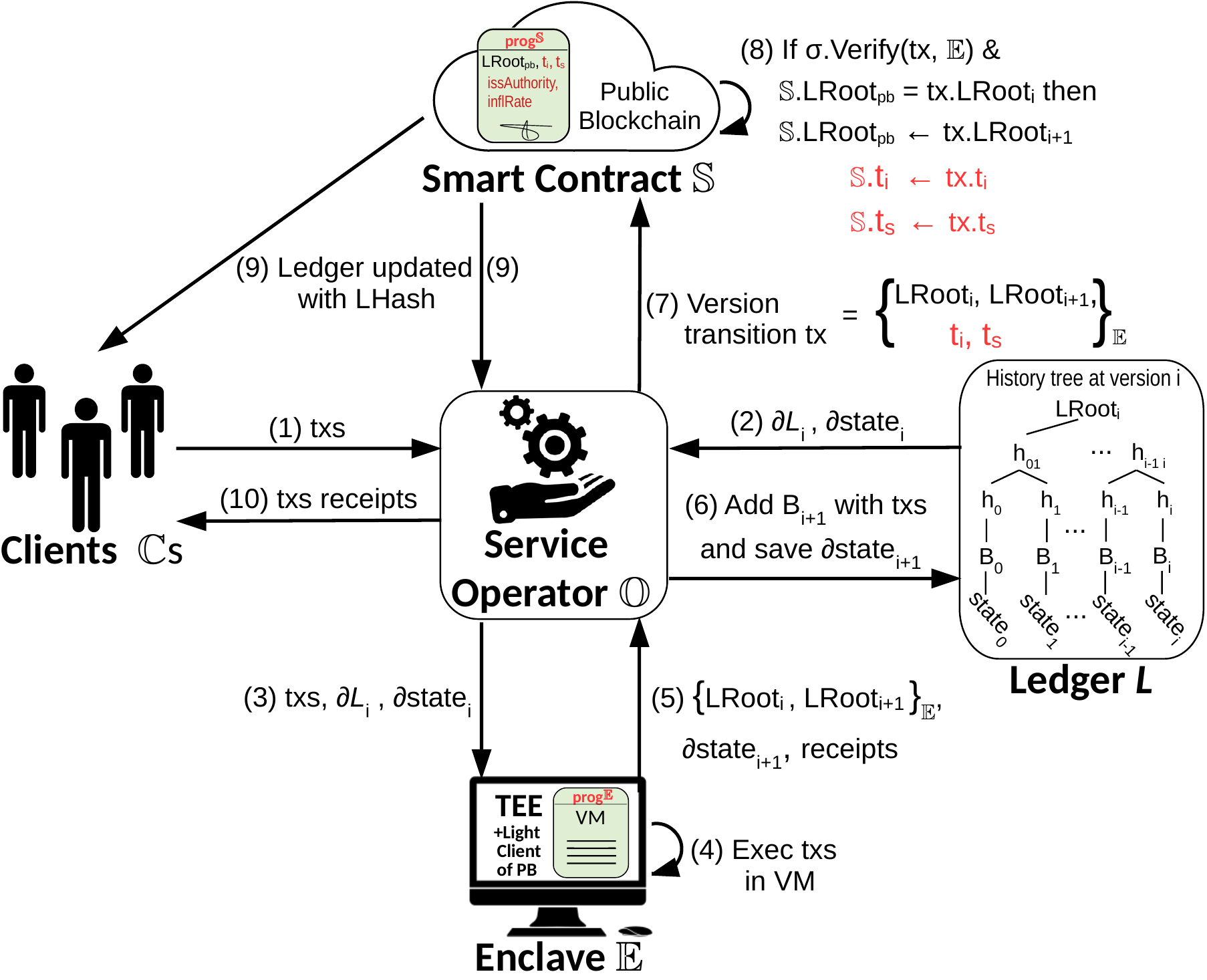}
	\caption{Architecture of Aquareum with our modifications in red.}
	\label{fig:architecture-aquareum}
	\vspace{-0.4cm}
\end{figure}
The overview of Aquareum is depicted in \autoref{fig:architecture-aquareum} (where parts in red are our modifications and are irrelevant for the current description). 
In Aquareum, clients $\mathbb{C}$s submit transactions to operator $\mathbb{O}$ (1), who executes them in protected TEE enclave $\mathbb{E}$ (4) upon fetching a few data of the ledger $L$ with the partial state containing only concerned accounts (2).
$\mathbb{E}$ outputs updated state of affected client accounts with execution receipts and \textit{a version transition pair} of $L$ (5) that is periodically submitted to the smart contract $\mathbb{S}$ deployed on a public blockchain (7).
$\mathbb{S}$ verifies $\mathbb{E}$'s signature and the consistency of the previous version of $L$ with $\mathbb{S}$'s local snapshot (8) before updating the snapshot to a new version.
Note that snapshot is represented by the root hash (i.e., $LRoot$ ) of the history tree of $L$.

 \section{Problem Definition}
 \label{sec:problem}
 Our goal is to propose a CBDC approach that respects the features proposed in DEA manifesto~\cite{cbdc-manifesto} released in 2022, while on top of it, we assume other features that might bring more benefits and guarantees.
First, we start with a specification of the desired features related to a single instance of CBDC that we assume is operated by a single entity (further a bank or its operator) that maintains its ledger. 
Later, we describe desired features related to multiple instances of CBDC that co-exist in the ecosystem of wholesale and/or retail CBDC.\footnote{Note that we will propose two deployment scenarios (see \autoref{sec:identity-management-of-CBDC}), one for the wholesale environment and the second one for the retail environment of multiple retail banks interacting with a single central bank.}
In both cases, we assume that a central bank might not be a trusted entity.
All features that respect this assumption are marked with asterisk $^*$ and are considered as requirements for such an attacker model.

\subsection{Single Instance of CBDC}\label{sec:problem-features-single}
When assuming a basic building block of CBDC -- a single bank's CBDC working in an isolated environment from the other banks -- we specify the desired features of CBDC as follows:
\begin{compactenum}

    \item[\textbf{Correctness of Operation Execution$^*$:}] 
    The clients who are involved in a monetary operation (such as a transfer) should be guaranteed with a correct execution of their operation.
    
    \item[\textbf{Integrity$^*$:}] 
    The effect of all executed operations made over the client accounts should be irreversible, and no ``quiet'' tampering of the data by a bank should be possible.
    Also, no conflicting transactions can be (executed and) stored by the CBDC instance in its ledger. 
    
    \item[\textbf{Verifiability$^*$:}] 
    This feature extends integrity and enables the clients of CBDC to obtain easily verifiable evidence that the ledger they interact with is internally correct and consistent.         
    In particular, it means that none of the previously inserted transactions was neither modified nor deleted. 
    
    \item[\textbf{Non-Equivocation$^*$:}]
    From the perspective of the client's security, the bank should not be able to present at least two inconsistent views on its ledger to (at least) two distinct clients who would accept such views as valid.
    
    \item[\textbf{Censorship Evidence$^*$:}]
    The bank should not be able to censor a client's request without leaving any public audit trails proving the censorship occurrence.
    
    \item[\textbf{Transparent Token Issuance$^*$:}] Every CBDC-issued token should be publicly visible (and thus audit-able) to ensure that a bank is not secretly creating token value ``out-of-nothing,'' and thus causing uncontrolled inflation.
    The transparency also holds for burning of existing tokens.

    \item[\textbf{High Performance:}] A CBDC instance should be capable of processing a huge number of transactions per second since it is intended for daily usage by thousands to millions of people.
    
    \item[\textbf{Privacy:}]
    All transfers between clients as well as information about the clients of CBDC should remain private for the public and all other clients that are not involved in particular transfers.
    However, a bank can access this kind of information and potentially provide it to legal bodies, if requested.
\end{compactenum}

\subsection{Multiple Instances of CBDC}
In the case of  multiple CBDC instances that can co-exist in a common environment, we extend the features described in the previous listing by features that are all requirements:
\begin{compactenum}
    
    \item[\textbf{Interoperability$^*$:}] 
    As a necessary prerequisite for co-existence of multiple CBDC instances, we require them to be mutually interoperable, which means that tokens issued by one bank can be transferred to any other bank.
    For simplicity, we assume that all the CBDC instances are using the unit token of the same value within its ecosystem.\footnote{On the other hand, conversions of disparate CBDC-backed tokens would be possible by following trusted oracles or oracle networks.}
    At the hearth of interoperability lies atomicity of supported operations.
    Atomic interoperability, however, requires means for accountable coping with censorship and recovery from stalling. 
    We specify these features in the following.
    
    \begin{compactenum}
        \item[\textbf{Atomicity$^*$:}] 
        Any operation (e.g., transfer) between two interoperable CBDC instances must be either executed completely or not executed at all. 
        As a consequence,
        no new tokens can be created out-of-nothing and no tokens can be lost in an inter-bank operation.
        Note that even if this would be possible, the state of both involved instances of CBDC would remain internally consistent; therefore, consistency of particular instances (\autoref{sec:problem-features-single}) is not a sufficient feature to ensure atomicity within multiple interoperable CBDC instances.
        This requirement is especially important due to trustless assumption about particular banks, who might act in their benefits even for the cost of imposing the extreme inflation to the whole system.\footnote{For example, if atomicity is not enforced, one bank might send the tokens to another bank, while not decreasing its supply due to pretended operation abortion.}
        
        \item[\textbf{Inter-CBDC Censorship Evidence$^*$:}] 
        Having multiple instances of CBDC enables a different way of censorship, where one CBDC (and its clients) might be censored within some inter-CBDC operation with another CBDC instance, precluding them to finish the operation.
        Therefore, there should exist a means how to accountably detect this kind of censorship as well.
    
        \item[\textbf{Inter-CBDC Censorship Recovery$^*$:}] 
        If the permanent censorship happens and is indisputably proven, it must not impact other instances of CBDC, including the ones that the inter-CBDC operations are undergoing.
        Therefore, the interoperable CBDC environment should provide a means to recover from inter-CBDC censorship of unfinished operations.
    \end{compactenum}

    \item[\textbf{Identity Management of CBDC Instances$^*$:}] 
    Since we assume that CBDC in\-stan\-ces are trustless, in theory, there might emerge a fake CBDC instance, pretentding to act as a valid one.
    To avoid this kind of situation, it is important for the ecosystem of wholesale CBDC to manage identities of particular valid CBDC instances in a secure manner.
\end{compactenum}

\subsection{Adversary Model}
The attacker can be represented by the operator of a bank or the client of a bank, and their intention is to break functionalities that are provided by the features described above. 
Next, we assume that the adversary cannot undermine the cryptographic primitives used, the blockchain platform, and
the TEE platform deployed.

 \section{Proposed CBDC Approach}
 \label{sec:design}

\begin{figure*}[h!]
	\includegraphics[width=0.7\textwidth]{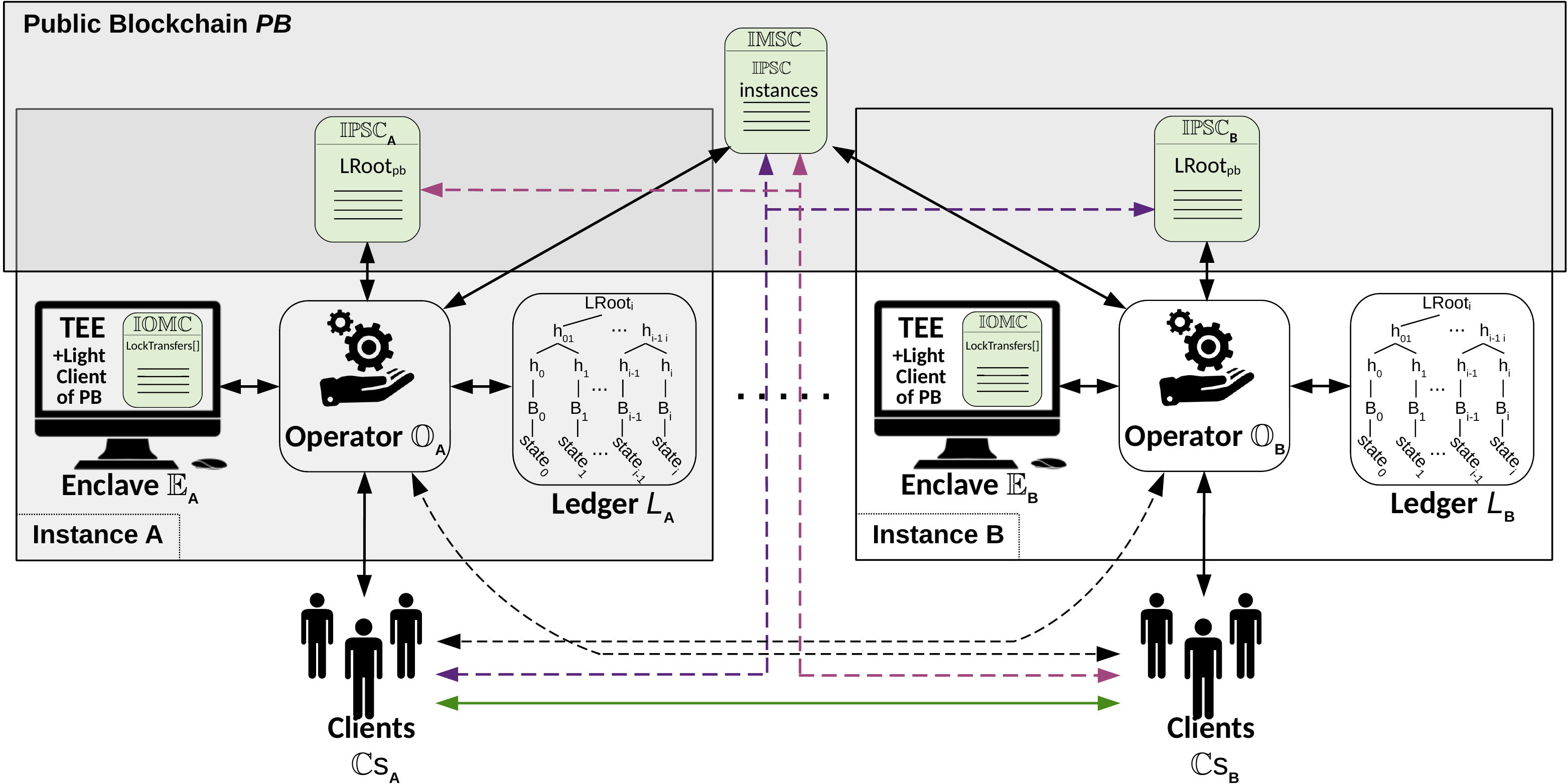}
	\vspace{-0.3cm}
	\caption{Overview of our CBDC architecture supporting interoperability among multiple CBDC instances (i.e., banks). The schema depicts two instances, where each of them has its own centralized ledger $L$ modified in a secure way through TEE of $\mathbb{E}$, while its integrity is ensured by periodic integrity snapshots to the integrity preserving smart contract ($\mathbb{IPSC}$) in a public blockchain $PB$. Each CBDC instance is registered in the identity management smart contract $\mathbb{IMSC}$ of a public blockchain, serving as a global registry of bank instances. A client who makes an inter-bank transfer communicates with her bank and the counter-party bank utilizing interoperability micro contracts ($\mathbb{IOMC}$), running in the TEE. Any censored request of a client is resolved by $\mathbb{IPSC}$ of a particular bank and can be initiated by its client or a counter-party client. 
}
    \label{fig:architecture}
\end{figure*}

We propose a holistic approach for the ecosystem of wholesale and/or retail CBDC, which aims at meeting the features described in \autoref{sec:problem}. 
To accomplish these features, we leverage interesting properties stemming from a combination of a public blockchain (with smart contract platform) and TEE. 
Such a combination was proposed for various purposes in related work (see \autoref{sec:related}), out of which the use case of generic centralized ledger Aquareum~\cite{homoliak2020aquareum} is most convenient to build on.
Therefore, we utilize Aquareum as a building block for a single instance of CBDC, 
and we make a few CBDC-specific modifications to it, enhancing its transparency and functionality.
Our modifications are outlined in \autoref{fig:architecture-aquareum} by red color, while the details of them (especially changes in programs of smart contract and enclave) will be described in this section.
First, we start by a description of a single CBDC instance and then we extend it to a fully interoperable environment consisting of multiple CBDC instances.

Note that in this paper, we focus solely on the transfer of tokens operation within the context of CBDC interoperability.
However, our approach could be extended to different operations, involving inter-CBDC smart contract invocations.
Also, note that to distinguish between smart contracts on a public blockchains and smart contracts running in TEE, we will denote latter as \textbf{micro contracts} (or $\mu$-contracts).
Similarly, we denote transactions sent to TEE as \textbf{micro transactions} (or $\mu$-transactions) and blocks created in the ledger of CBDC instance as \textbf{micro blocks} (or $\mu$-blocks).

\subsection{\textbf{A CBDC Instance}}
Alike in Aquareum, the primary entity of each CBDC instance is its operator $\mathbb{O}$ (i.e., a bank), who is responsible for (1) maintaining the ledger $L$, 
(2) running the TEE enclave $\mathbb{E}$, 
(3) synchronization of the $L$'s snapshot to a public blockchain with smart contract $\mathbb{IPSC}$ ($\mathbb{I}$ntegrity $\mathbb{P}$reserving $\mathbb{S}$mart $\mathbb{C}$ontract), 
(4) resolving censorship requests, and 
(5) a communication with clients $\mathbb{C}$s.

\subsubsection{\textbf{Token Issuance}}
On top of Aquareum's $\mathbb{S}$, our $\mathbb{IPSC}$ contains snapshotting of the total issued tokens $t_i$ by the current CBDC instance and the total supply $t_s$ available at the instance for the purpose of transparency in token issuance (and potentially even burning).
Therefore, we extend the $\mathbb{E}$-signed version transition pair periodically submitted to $\mathbb{IPSC}$ by these two fields that are relayed to $\mathbb{IPSC}$ upon snapshotting $L$ (see red text in \autoref{fig:architecture-aquareum}).
Notice that $t_i = t_s$ in the case of a single instance since the environment of the instance is isolated.

\subsubsection*{\textbf{An Inflation Bound}}
Although snapshotting the total tokens in circulation is useful for the transparency of token issuance, $\mathbb{O}$ might still hyper-inflate the CBDC instance.
Therefore, we require $\mathbb{O}$ to guarantee a maximal inflation rate $i_r$ per year, which can be enforced by $\mathbb{IPSC}$ as well as $\mathbb{E}$ since the code of both is publicly visible and attestable.
The $i_r$ should be adjusted to a constant value by $\mathbb{O}$ at the initialization of $\mathbb{IPSC}$ and verified every time the new version of $L$ is posted to $\mathbb{IPSC}$; 
in the case of not meeting the constrain, the new version would not be accepted at $\mathbb{IPSC}$.
However, another possible option is that the majority vote of $\mathbb{C}$s can change $i_r$ even after initialization.
Besides, $\mathbb{E}$ also enforces $i_r$ on $t_i$ and does not allow $\mathbb{O}$ to issue yearly more tokens than defined by $i_r$. 
Nevertheless, we put the inflation rate logic also into $\mathbb{IPSC}$ for the purpose of transparency.

\subsubsection{\textbf{Initialization}}
First, $\mathbb{E}$ with program $prog^{\mathbb{E}}$ (see  \autoref{alg:enclave-VM} of Appendix) generates and stores two key pairs, one under $\Sigma_{pb}$ (i.e., $SK_{\mathbb{E}}^{pb}$, $PK_{\mathbb{E}}^{pb}$) and one under $\Sigma_{tee}$ (i.e.,  $SK_{\mathbb{E}}^{tee}$, $PK_{\mathbb{E}}^{tee}$).
Then, $\mathbb{O}$ generates one key pair under $\Sigma_{pb}$ (i.e.,  $SK_{\mathbb{O}}^{pb}$, $PK_{\mathbb{O}}^{pb}$), which is then  used as the sender of a transaction deploying $\mathbb{IPSC}$ with program $prog^{\mathbb{IPSC}}$ (see \autoref{alg:IPSC-smart-contract} of Appendix) at public blockchain with parameters $PK_{\mathbb{E}}^{pb}$, $PK_{\mathbb{E}}^{tee}$, $PK_{\mathbb{O}}^{pb}$, $t_i$, and $i_r$.
Then, $\mathbb{IPSC}$ stores the keys in parameters, sets the initial version of $L$ by putting $LRoot_{pb} \leftarrow ~\perp$, and sets the initial total issued tokens and the total supply, both to $t_i$.\footnote{Among these parameters, a constructor of $\mathbb{IPSC}$ also accepts the indication whether an instance is allowed to issue tokens. This is, however, implicit for the single instance, while restrictions are reasonable in the case of multiple instances.}

\subsubsection*{\textbf{Client Registration}}
A client $\mathbb{C}$ registers with $\mathbb{O}$, who performs know your customer (KYC) checks and submits her public key $PK_{pb}^{\mathbb{C}}$ to $\mathbb{E}$.
Then, $\mathbb{E}$ outputs an execution receipt about the successful registration of $\mathbb{C}$ as well as her access ticket $t^{\mathbb{C}}$ that will serve for potential communication with $\mathbb{IPSC}$ and its purpose is to avoid spamming $\mathbb{IPSC}$ by invalid requests. 
In detail, $t^{\mathbb{C}}$ is the $\mathbb{E}$-signed tuple that contains $PK_{pb}^{\mathbb{C}}$ and optionally other fields such as the account expiration timestamp. 
Next, $\mathbb{C}$ verifies whether her registration (proved by the receipt) was already snapshotted by $\mathbb{O}$ at $\mathbb{IPSC}$.

\subsubsection{\textbf{Normal Operation}}
$\mathbb{C}$s send \textit{$\mu$-transactions} (writing to $L$) and \textit{queries} (reading from $L$) to $\mathbb{O}$, who validates them and relays them to $\mathbb{E}$, which processes them within its virtual machine (Aquareum uses eEVM~\cite{eEVM-Microsoft}).
Therefore, $L$ and its state are modified in a trusted code of $\mathbb{E}$, creating a new version of $L$, which is represented by the root hash $LRoot$ of the history tree.
Note that program $prog^{\mathbb{E}}$ is public and can be remotely attested by $\mathbb{C}$s (or anybody).
$\mathbb{O}$ is responsible for a periodic synchronization of the most recent root hash $LRoot_{cur}$ (i.e., snapshotting the current version of $L$ ) to $\mathbb{IPSC}$, running on a public blockchain $PB$. 
Besides, $\mathbb{C}$s use this smart contract to resolve censored transactions and queries, while preserving the privacy of data. 

\subsubsection{\textbf{Censorship Resolution}}\label{sec:approch-single-instance-cens}
$\mathbb{O}$ might potentially censor some \textit{write} transactions or \textit{read} queries of $\mathbb{C}s$.
However, these can be resolved by Aquareum's mechanism as follows. 
If $\mathbb{C}$'s $\mu$-transaction $\mu$-tx is censored by $\mathbb{O}$, $\mathbb{C}$ first creates $PK_{\mathbb{E}}^{tee}$-encrypted $\mu$-$etx$  (to ensure privacy in $PB$), and then she creates and signs a transaction containing $\mathbb{C}'s$ access ticket $t^{\mathbb{C}}$ and $\mu$-$etx$.
$\mathbb{C}$ sends this transaction to $\mathbb{IPSC}$, which verifies $t^{\mathbb{C}}$ and stores $\mu$-$etx$, which is now visible to $\mathbb{O}$ and the public.
Therefore, $\mathbb{O}$ might relay $\mu$-$etx$ to $\mathbb{E}$ for processing and then provide $\mathbb{E}$-signed execution receipt to $\mathbb{IPSC}$ that publicly resolves this censorship request.
On the other hand, if $\mathbb{O}$ were not to do it, $\mathbb{IPSC}$ would contain an indisputable proof of censorship by $\mathbb{O}$ on a client $\mathbb{C}$.

\subsection{\textbf{Multiple CBDC Instances}}
The conceptual model of our interoperable CBDC architecture is depicted in \autoref{fig:architecture}.
It consists of multiple CBDC instances (i.e., at least two), whose $\mathbb{C}s$ communicate in three different ways: (1) directly with each other, (2) in the instance-to-instance fashion through the infrastructure of their $\mathbb{O}$ as well as counterpart's $\mathbb{O}$, (3) through $PB$ with $\mathbb{IPSC}$ of both $\mathbb{O}$s and a global registry $\mathbb{IMSC}$ managing identities of instances. 

For simplified description, in the following we assume the transfer operation where a local CBDC instance in \autoref{fig:architecture} is A (i.e., the sender of tokens) and the external one is B (i.e., the receiver of tokens).
To ensure interoperability, we require a communication channel of local clients $\mathbb{C}s_A$ to external clients $\mathbb{C}s_{B}$ (the green arrow), the local operator $\mathbb{O}_{A}$ (the black arrow), and the external operator $\mathbb{O}_{B}$  (the black dashed arrow).
In our interoperability protocol $\Pi^{T}$ (described later in \autoref{sec:our-protocol}), external $\mathbb{C}s_{B}$ use the channel with the local operator $\mathbb{O}_{A}$ only for obtaining incremental proofs of $L_A$'s history tree to verify inclusion of some $\mu$-transactions in $L_A$.
However, there might arise a situation in which $\mathbb{O}_{A}$ might censor such queries, therefore, we need to address it by another communication channel -- i.e., the public blockchain $PB$.

\subsubsection*{\textbf{Censorship of External Clients}}
We allow external clients $\mathbb{C}s_{B}$ to use the same means of censorship resolution as internal clients of a single CBDC instance (see \autoref{sec:approch-single-instance-cens}). 
To request a resolution of a censored query, the external $\mathbb{C}_B$ uses the access ticket $t^{\mathbb{C}_B}$ at $\mathbb{IPSC}_A$, which is issued by $\mathbb{E}_A$ in the first phase of $\Pi^{T}$. 

\subsubsection*{\textbf{Identification of Client Accounts}}
To uniquely identify $\mathbb{C}$'s account at a particular CBDC instance, first it is necessary to specify the globally unique identifier of the CBDC instance. 
The best candidate is the blockchain address of the $\mathbb{IPSC}$ in $PB$ since it is publicly visible and unique in $PB$ (and we denote it by $\mathbb{IPSC}$). 
Then, the identification of $\mathbb{C}$'s relevant account is a pair $\mathbb{C}^{ID} = \{PK_{pb}^{\mathbb{C}}||~ \mathbb{IPSC}\}$.
Note that $\mathbb{C}$ might use the same $PK_{pb}^{\mathbb{C}}$ for the registration at multiple CBDC instances (i.e., equivalent of having accounts in multiple banks); however, to preserve better privacy, making linkage of $\mathbb{C}$'s instances more difficult, we recommend $\mathbb{C}s$ to have dedicated key pair for each instance.

\subsubsection{\textbf{Identity Management of CBDC Instances}}\label{sec:identity-management-of-CBDC}
To manage identities of all CBDC instances in the system, we need a global registry of their identifiers -- $\mathbb{IPSC}$ addresses.
For this purpose, we use the $\mathbb{I}$dentity $\mathbb{M}$anagement $\mathbb{S}$mart $\mathbb{C}$ontract ($\mathbb{IMSC}$) deployed in $PB$ (see program $prog^{\mathbb{IMSC}}$ in \autoref{alg:imsc}).
We propose $\mathbb{IMSC}$ to be managed in either  decentralized or centralized fashion, depending on the deployment scenario described below. 

\subsection*{\textbf{Deployment scenarios}}

\subsubsection*{\textbf{Decentralized Scheme}}
In the decentralized scheme, the enrollment of a new CBDC instance must be approved by a majority vote of the already existing instances.
This might be convenient for interconnecting  central banks from various countries/regions.

The enrollment requires creating a request entry at $\mathbb{IMSC}$ (i.e., $newJoinRequest()$) by a new instance specifying the address of its $\mathbb{IPSC}_{new}$ and $PK_{PB}^{\mathbb{O}_{new}}$.
Then, the request has to be approved by voting of existing instances. 
Prior to voting (i.e., $approveJoinRequest()$), the existing instances should first verify a new instance by certain legal processes as well as by technical means: 
do the remote attestation of $prog^\mathbb{E}_{new}$, verify the inflation rate $i_r$ and the initial value of total issued tokens $t_i$ in $\mathbb{IPSC}$, etc.
Removing of the existing instance also requires the majority of all instances, who should verify legal conditions prior to voting.

\subsubsection*{\textbf{Centralized Scheme}} 
So far, we were assuming that CBDC instances are equal, which might be convenient for interconnection of central banks from different countries.
However, from the single-country point-of-view, there usually exist only one central bank, which might not be interested in decentralization of its competences (e.g., issuing tokens, setting inflation rate) among multiple commercial banks. 
We respect this and enable our approach to be utilized for such a use case, while the necessary changes are made to $\mathbb{IMSC}_c$ (see \autoref{alg:imsc-centralized}), allowing to have only one CBDC authority that can add or delete instances of (commercial) banks, upon their verification (as outlined above).
The new instances can be adjusted even with token issuance capability and constraints on inflation, which is enforced within the code of $\mathbb{E}$ as well as $\mathbb{IPSC}$.
\begin{algorithm}[t] 
	\scriptsize 
	\caption{$prog^{\mathbb{IMSC}}_d$ of decentralized $\mathbb{IMSC}$ }\label{alg:imsc}
	
	\SetKwProg{func}{function}{}{}
	
	\smallskip
	$\triangleright$ \textsc{Declaration of types and variables:}\\
	\hspace{1em} $msg$: a current transaction that called $\mathbb{IMSC}$,  \\
	\hspace{1em} struct \textbf{InstanceInfo} \{ \\
	\begin{scriptsize}
		\hspace{3em} $operator$ : $PK_{\mathbb{O}}^{PB}$ of the instance's $\mathbb{O}$,\\
		\hspace{3em} $isApproved$: admission status of the instance,\\
		\hspace{3em} $approvals \leftarrow []$ : $\mathbb{O}$s who have approved the instance creation (or deletion),\\	
	\end{scriptsize}
	\hspace{1em} \} \\
	
	\hspace{1em} $instances[]$: a mapping of $\mathbb{IPSC}$ to \textbf{InstanceInfo}, \\
	\smallskip
	$\triangleright$ \textsc{Declaration of functions:}\\
	\func{$Init$($\mathbb{IPSC}s[], \mathbb{O}s[]$) \textbf{public} \Comment{Initial instances are implicitly approved.} }{
		\textbf{assert} $|\mathbb{IPSC}s| = |\mathbb{O}s|$ ; \\
		
		\For{$i \leftarrow 0;\ i \le |\mathbb{O}s|;\ i \leftarrow i + 1$}{ 
			
			$instances[\mathbb{IPSC}s[i]] \leftarrow \textbf{InstanceInfo}(\mathbb{O}s[i], True, [])$; \\
		}
	}
	
	\smallskip
	\func{$newJoinRequest$($\mathbb{IPSC}$) \textbf{public} }{
		\textbf{assert} $instances[\mathbb{IPSC}] = ~\perp$;  \Comment{The instance must not exist yet.}\\
		$instances[\mathbb{IPSC}] \leftarrow \textbf{InstanceInfo}(msg.sender, False, [])$; \\
	}
	\smallskip
	\func{$approveJoinRequest$($\mathbb{IPSC}_{my}, \mathbb{IPSC}_{new}$) \textbf{public} }{
		\textbf{assert} $instances[\mathbb{IPSC}_{my}].operator = msg.sender$; \Comment{Sender's check.}\\
		\textbf{assert} $instances[\mathbb{IPSC}_{my}].isApproved$; \Comment{The sending $\mathbb{O}$ has valid instance.}\\
		\textbf{assert} $!instances[\mathbb{IPSC}_{new}].isApproved$; \Comment{The new instance is not approved.}\\
		
		$r \leftarrow instances[\mathbb{IPSC}_{new}]$; \\
		$r.approvals[msg.sender] \leftarrow True$; \Comment{The sender acknowledges the request.}\\
		\If{$|r.approvals| > \lfloor |instances| / 2 \rfloor$}{
			$r.isApproved \leftarrow True$; \Comment{Majority vote applies.}\\
			$r.approvals \leftarrow []$; \Comment{Switch this field for a potential deletion.} \\
		} 
	}
	\smallskip
	\func{$approveDelete$($\mathbb{IPSC}_{my}, \mathbb{IPSC}_{del}$) \textbf{public} }{
		\textbf{assert} $instances[\mathbb{IPSC}_{my}].operator = msg.sender$; \Comment{Sender's check.}\\
		\textbf{assert} $instances[\mathbb{IPSC}_{my}].isApproved$; \Comment{The sending $\mathbb{O}$ has valid instance.}\\
		\textbf{assert} $instances[\mathbb{IPSC}_{del}].isApproved$; \Comment{An instance to delete must be approved.}\\
		
		$r \leftarrow instances[\mathbb{IPSC}_{del}]$; \\
		$r.approvals[msg.sender] \leftarrow True$; \Comment{The sender acknowledges the request.}\\
		\If{$|r.approvals| > \lfloor |instances| / 2 \rfloor$}{
			\textbf{delete} $r$;
		} 
	}
	
	\smallskip

	\vspace{-0.1cm}
\end{algorithm}

\begin{algorithm}[t] 
	\scriptsize 
	\caption{$prog^{\mathbb{IMSC}}_c$ of centralized $\mathbb{IMSC}$ }\label{alg:imsc-centralized}
	
	\SetKwProg{func}{function}{}{}
	
	\smallskip
	$\triangleright$ \textsc{Declaration of types and variables:}\\
	\hspace{1em} $msg$: a current transaction that called $\mathbb{IMSC}$,  \\
	\hspace{1em} $authority$: $\mathbb{IPSC}$ of the authority bank, \\
	\hspace{1em} $authority^\mathbb{O}$: $PK^\mathbb{O}_{pb}$ of $\mathbb{O}$ at authority bank, \\
	\hspace{1em} $instances[]$: a mapping of $\mathbb{IPSC}$ to $PK_{\mathbb{O}}^{PB}$, \\
	\smallskip
	$\triangleright$ \textsc{Declaration of functions:}\\
	\func{$Init$($\mathbb{IPSC}$) \textbf{public} \Comment{Initial instances are implicitly approved.} }{				
		$authority^\mathbb{O} \leftarrow msg.sender$; \\
		$authority \leftarrow \mathbb{IPSC}$; \\
	}
	
	\smallskip
	\func{$addInstance$($\mathbb{IPSC}_{new}, ~\mathbb{O}_{new}$) \textbf{public} }{
		\textbf{assert} $msg.sender =  authority^\mathbb{O}$ ; \Comment{Only the authority can add instances.}\\
		\textbf{assert} $instances[\mathbb{IPSC}_{new}] = ~\perp$;  \Comment{The instance must not exist yet.}\\
		$instances[\mathbb{IPSC}_{new}] \leftarrow \mathbb{O}_{new}$; \\
	}
	\smallskip
	
	\func{$delInstance$($\mathbb{IPSC}_{del}$) \textbf{public} }{
		\textbf{assert} $msg.sender =  authority^\mathbb{O}$ ; \Comment{Only the authority can delete instances.}\\
		\textbf{delete} $instances[\mathbb{IPSC}_{del}]$; \\
	}
	\smallskip
	\vspace{-0.1cm}
\end{algorithm}

\subsubsection{\textbf{Token Issuance}}
With multiple CBDC instances, $\mathbb{C}$s and the public can obtain the total value of issued tokens in the ecosystem of CBDC and compare it to the total value of token supply of all instances.
Nevertheless, assuming only two instances A and B, the value of $t_s$ snapshotted by $\mathbb{IPSC}_A$ might not reflect the recently executed transfers to instance B that might have already made the snapshot of its actual $L_B$ version to $\mathbb{IPSC}_B$, accounting for the transfers.
As a consequence, given a set of instances, the value of the aggregated $t_s$ should always be greater or equal than the corresponding sum of $t_i$:
\begin{eqnarray}
	t_i^A + t_i^B  &\leq& t_s^A + t_s^B. 
\end{eqnarray}
We can generalize it for $N$ instances known by $\mathbb{IMSC}$ as follows:
\begin{eqnarray}\label{eq:token-sum-multi}
	\sum_{\forall X ~\in~ \mathbb{IMSC}} t_i^X   &\leq& \sum_{\forall X ~\in~ \mathbb{IMSC}} t_s^X. 
\end{eqnarray}

\subsubsection{\textbf{Inflation Rate}}
In contrast to a single CBDC instance, multiple independent instances must provide certain guarantees about inflation not only to their clients, but also to each other.
For this purpose, the parameter inflation rate $i_r$ is adjusted to a constant value in the initialization of $\mathbb{IPSC}$ and checked before the instance is approved at $\mathbb{IMSC}$.

If one would like to enable the update of $i_r$ at CBDC instances, a majority vote at $\mathbb{IMSC}$ on a new value could be utilized (or just the vote of authority in the case of centralized scenario). 
Nevertheless, to support even fairer properties, $\mathbb{C}$s of a particular instance might vote on the value of $i_r$ upon its acceptance by $\mathbb{IOMC}$ and before it is propagated to $\mathbb{IPSC}$ of an instance.
Then, based on the new value of $\mathbb{IPSC}.i_r$, $\mathbb{E}.i_r$  can be adjusted as well (i.e., upon the validation by the light client of $\mathbb{E}$). 
However, the application of such a mechanism might depend on the use case, and we state it only as a possible option that can be enabled in our approach.  

\begin{algorithm}[t] 
	\caption{$prog^{\mathbb{IOMC}^{S}}$ of sending $\mathbb{IOMC}^{S}$ }\label{alg:iomc-send}
	\scriptsize
	\SetKwProg{func}{function}{}{}
	
	\smallskip
	$\triangleright$ \textsc{Declaration of types and variables:}\\
	
	\hspace{1em} $\mathbb{E}$, \Comment{The reference to $\mathbb{E}_A$ of sending party. } \\
	\hspace{1em} $msg$, \Comment{The current $\mu$-transaction that called $\mathbb{IOMC}^S$.}  \\
	\hspace{1em} struct \textbf{LockedTransfer} \{ \\
	\hspace{3em} $sender$, \Comment{Sending client $\mathbb{C}_A$.} \\
	\hspace{3em} $receiver$, \Comment{Receiving client $\mathbb{C}_B$.} \\
	\hspace{3em} $receiver\mathbb{IPSC}$, \Comment{The $\mathbb{IPSC}$ contract address of the receiver's instance.}\\
	\hspace{3em} $amount$, \Comment{Amount of tokens sent.}\\
	\hspace{3em} $hashlock$,  \Comment{Hash of the secret of the sending $\mathbb{C}_A$.}\\
	\hspace{3em} $timelock$, \Comment{A timestamp defining the end of validity of the transfer.}\\
	\hspace{3em} $isCompleted$, \Comment{Indicates whether the transfer has been completed.}\\
	\hspace{3em} $isReverted$, \Comment{Indicates whether the transfer has been canceled.}\\ 
	\hspace{1em} \}, \\

	\hspace{1em} $transfers \leftarrow  []$, \Comment{Initiated outgoing transfers (i.e., LockedTransfer).}  \\
	\hspace{1em}\textbf{const} $timeout^{HTLC} \leftarrow 24h$, \Comment{Set the time lock for e.g., 24 hours.}   \\
	
	\smallskip
	$\triangleright$ \textsc{Declaration of functions:}\\
	
	\func{$sendInit$($receiver, receiver\mathbb{IPSC}, hashlock$) \textbf{public payable} }{
		\textbf{assert} $msg.value > 0$; \Comment{Checks the amount of tokens.} \\
		$timelock \leftarrow timestamp.now() + timeout^{HTLC}$; \\
		$t \leftarrow \textbf{LockedTransfer}(msg.sender, receiver, receiver\mathbb{IPSC}, $\\
		\hspace{1em} $msg.value$, $hashlock, timelock, False, False)$; \Comment{A new receiving transfer.} \\
		$transfers.append(t)$; \\
		
		\textbf{Output} $("sendInitialized", transferID \leftarrow |transfers| - 1))$; \\
	}
	\smallskip
	\func{$sendCommit$($transferID, secret, extTransferID$) \textbf{public} }{
		\textbf{assert} $transfers[transferID] \neq \perp$;  \Comment{Check the existence of locked transfer.}\\
		$t \leftarrow transfers[transferID]$; \\
		\textbf{assert} $t.hashlock = h(secret)$;  \Comment{Check the secret.}\\
		\textbf{assert} $!t.isCompleted ~\wedge~ !t.isReverted$;  \Comment{Test if the transfer is still pending.}\\
		$t.isCompleted \leftarrow True$; \\
		\textbf{burn} t.amount; \Comment{Burn tokens.}\\
		$\mathbb{E}.t_s \leftarrow \mathbb{E}.t_s - t.amount$; \Comment{Decrease the total supply of the instance.} \\
		\textbf{Output} $("sendCommitted",transferID,$ $extTransferID,$ $t.receiver,$ $t.receiver\mathbb{IPSC},$ $t.amount)$;\\
	}
	\smallskip
	\func{$sendRevert$($transferID$) \textbf{public} }{
		\textbf{assert} $transfers[transferID] \neq \perp$;  \Comment{Check the existence of locked transfer.}\\
		$t \leftarrow transfers[transferID]$; \\
		\textbf{assert} $!t.isCompleted ~\wedge~ !t.isReverted$;  \Comment{Test the transfer is still pending.}\\
		\textbf{assert} $t.timelock \leq timestamp.now()$;  \Comment{Check the HTLC expiration.}\\
		$transfer(t.amount, t.sender)$; \Comment{Returning tokens back to the sender.}\\
		$t.isReverted \leftarrow True$; \\
		\textbf{Output}$("sendReverted", transferID)$;\\
	}
	\smallskip
	\vspace{-0.1cm}
\end{algorithm}

\subsubsection{\textbf{Interoperability}}
The interoperability logic itself is provided by our protocol $\Pi^{T}$ that utilizes $\mathbb{I}$nter$\mathbb{O}$perability $\mathbb{M}$icro  $\mathbb{C}$on\-tracts $\mathbb{IOMC}^S$ and $\mathbb{IOMC}^R$, which serve for sending and receiving tokens, respectively.
Therefore, in the context of $\mathbb{E}$-isolated environment these $\mu$-contracts allow to mint and burn tokens, reflecting the changes in $t_s$ after sending or receiving tokens between CBDC instances.
Both $\mu$-contracts are deployed in $\mathbb{E}$ by each $\mathbb{O}$ as soon as the instance is created, while $\mathbb{E}$ records their addresses that can be obtained and attested by $\mathbb{C}$s.
We briefly review these contracts in the following, and we will demonstrate their usage in \autoref{sec:our-protocol}.

\subsubsection*{\textbf{The Sending $\mathbb{IOMC}^S$}}
The sending $\mathbb{IOMC}^S$ (see \autoref{alg:iomc-send}) is based on Hash Time LoCks (HTLC), thus upon initialization of transfer by $hashlock$ provided by $\mathbb{C}_A$ (i.e., $hashlock \leftarrow h(secret)$) and calling $sendInit(hash\-lock, \ldots)$,  $\mathbb{IOMC}^S$ locks transferred tokens for the timeout required to complete the transfer by $send\-Commit\-(secret, \ldots)$. 
If tokens are not successfully transferred to the recipient of the external instance during the timeout, they can be recovered by the sender (i.e., $sendRevert()$).\footnote{Note that setting a short timeout might prevent the completion of the protocol.}
If tokens were sent successfully from $\mathbb{C}_A$ to  $\mathbb{C}_B$, then instance A burns them within $sendCommit()$ of $\mathbb{IOMC}^S$ and deducts them from $t_s$.
Note that deducting $t_s$ is a special operation that cannot be executed within standard $\mu$-contracts, but $\mathbb{IOMC}$ contracts are exceptions and can access some variables of~$\mathbb{E}$.

\subsubsection*{\textbf{The Receiving $\mathbb{IOMC}^R$}}
\begin{figure}[b]
    \centering
    \vspace{-0.3cm}
	\includegraphics[width=0.9\columnwidth]{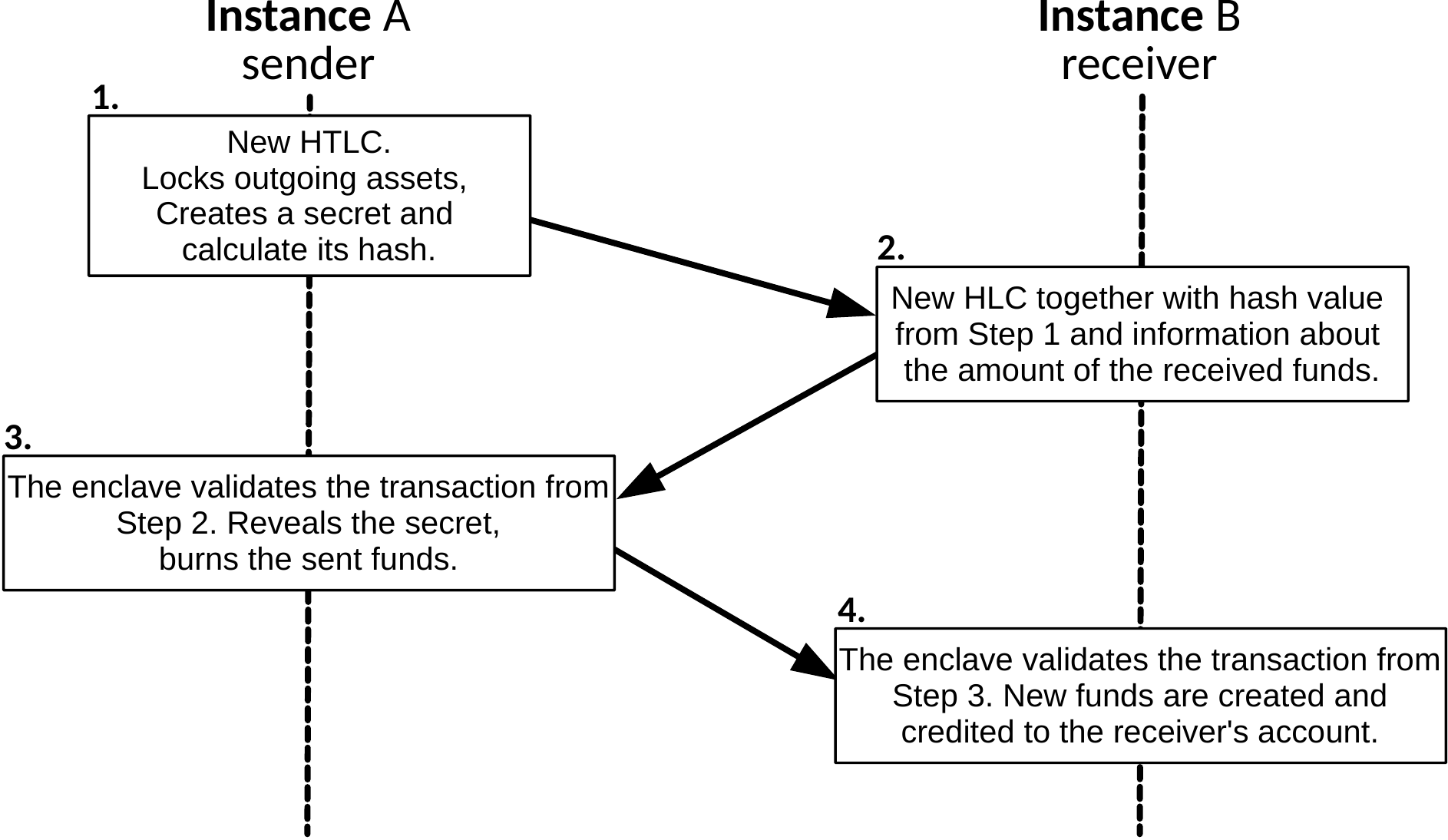}
	\vspace{-0.3cm}
	\caption{Overview of the protocol $\Pi^{T}$, consisting of 4 phases. 
		}
	\label{fig:protocol-simplified}
\end{figure}

The receiving  $\mathbb{IOMC}^R$ (see \autoref{alg:iomc-recv}) is based on Hashlocks (referred to as HLC) and works pairwise with sending $\mathbb{IOMC}^S$ to facilitate four phases of our interoperable transfer protocol $\Pi_{T}$ (described below).
After calling $\mathbb{IOMC^S}.sendInit()$, incoming initiated transfer is recorded at $\mathbb{IOMC}^R$ by $receiveInit(hash\-lock, \ldots)$.
Similarly, after executing token deduction at instance A (i.e., $\mathbb{IOMC^S}.send\-Commit\-(secret, \ldots)$), incoming transfer is executed at $\mathbb{IOMC}^R$ by $receiveCommit(secret, \ldots)$ that mints tokens to $\mathbb{C}_B$ and increases $t_s$.
Similar to $\mathbb{IOMC}^S$, minting tokens and increasing $t_s$ are special operations requiring access to $\mathbb{E}$, which is exceptional for $\mathbb{IOMC}$.
The overview of $\Pi_T$ is depicted in \autoref{fig:protocol-simplified}.

\begin{algorithm}[t] 
	\caption{$prog^{\mathbb{IOMC}^{R}}$ of receiving $\mathbb{IOMC}^{R}$}\label{alg:iomc-recv}
	\scriptsize
	\SetKwProg{func}{function}{}{}
	
	\smallskip
	$\triangleright$ \textsc{Declaration of types and variables:}\\
	\hspace{1em} $\mathbb{E}$, \Comment{The reference to $\mathbb{E}_B$ of receiving party.} \\
	
	\hspace{1em} struct \textbf{LockedTransfer} \{ \\
	\begin{scriptsize}
		\hspace{3em} $sender$, \Comment{Sending client $\mathbb{C}_A$.} \\
		\hspace{3em} $sender\mathbb{IPSC}$, \Comment{The IPSC contract address of the sender's instance.}\\
		\hspace{3em} $receiver$,\Comment{Receiving client $\mathbb{C}_B$.}  \\
		\hspace{3em} $amount$, \Comment{Amount of transferred tokens.}\\
		\hspace{3em} $hashlock$, \Comment{Hash of the secret of the sending $\mathbb{C}_A$.}\\
		\hspace{3em} $isCompleted$, \Comment{Indicates whether the transfer has been completed.}\\    	
	\end{scriptsize}
	\hspace{1em}	\}, \\
	
	\hspace{1em} $transfers \leftarrow  []$, \Comment{Initiated incoming transfers (i.e., LockedTransfer).}  \\	
	\smallskip
	$\triangleright$ \textsc{Declaration of functions:}
	
	\smallskip
	\func{$receiveInit$($sender, sender\mathbb{IPSC}, hashlock, amount$) \textbf{public} }{
		\textbf{assert} $amount > 0$;  \\
		$t \leftarrow \textbf{LockedTransfer}(sender, sender\mathbb{IPSC}, msg.sender, amount,$\\
		\hspace{6em} $hashlock, False)$; \Comment{Make a new receiving transfer entry.} \\
		$transfers.append(t)$; \\
		\textbf{Output}$("receiveInitialized", transferID \leftarrow |transfers| - 1)$;\\
	}
	\smallskip
	\func{$receiveCommit$($transferID, secret$) \textbf{public} }{
		\textbf{assert} $transfers[transferID] ~\neq~ \perp$;  \Comment{Check the existence of transfer entry.}\\
		$t \leftarrow transfers[transferID]$; \\
		\textbf{assert} $t.hashlock = h(secret)$;  \Comment{Check the secret.}\\
		\textbf{assert} $!t.isCompleted$;  \Comment{Check whether the transfer is pending.}\\
		
		$\mathbb{E}.mint(this, t.amount)$; \Comment{Call $\mathbb{E}$ to mint tokens on $\mathbb{IOMC}_R$.} \\
		$\mathbb{E}.t_s \leftarrow \mathbb{E}.t_s + t.amount$ ; \Comment{Increase the total supply of the instance.} \\
		$transfer(t.amount, t.receiver)$; \Comment{Credit tokens to the recipient.}\\
		$t.isCompleted \leftarrow True$; \\        
		\textbf{Output}$("receiveCommited", transferID)$;\\
		
	}
	\smallskip
	\vspace{-0.1cm}
\end{algorithm}

\subsection{Interoperable Transfer Protocol $\mathbf{\Pi^T}$}\label{sec:our-protocol}
In this section we outline our instance-to-instance interoperable transfer protocol $\Pi^{T}$ for inter-CBDC transfer operation, which is inspired by the atomic swap protocol (see \autoref{sec:atomicswap}), but in contrast to the exchange-oriented approach of atomic swap, $\Pi^{T}$ focuses only on one-way atomic transfer between instances of the custodial environment of CBDC, where four parties are involved in each transfer -- a sending $\mathbb{C}_A$ and $\mathbb{O}_A$ versus a receiving $\mathbb{C}_B$ and $\mathbb{O}_B$.
The goal of $\Pi^{T}$ is to eliminate any dishonest behavior by $\mathbb{C}s$ or $\mathbb{O}$s that would incur token duplication or the loss of tokens.

	\begin{figure*}[th]
		\includegraphics[width=0.95\textwidth]{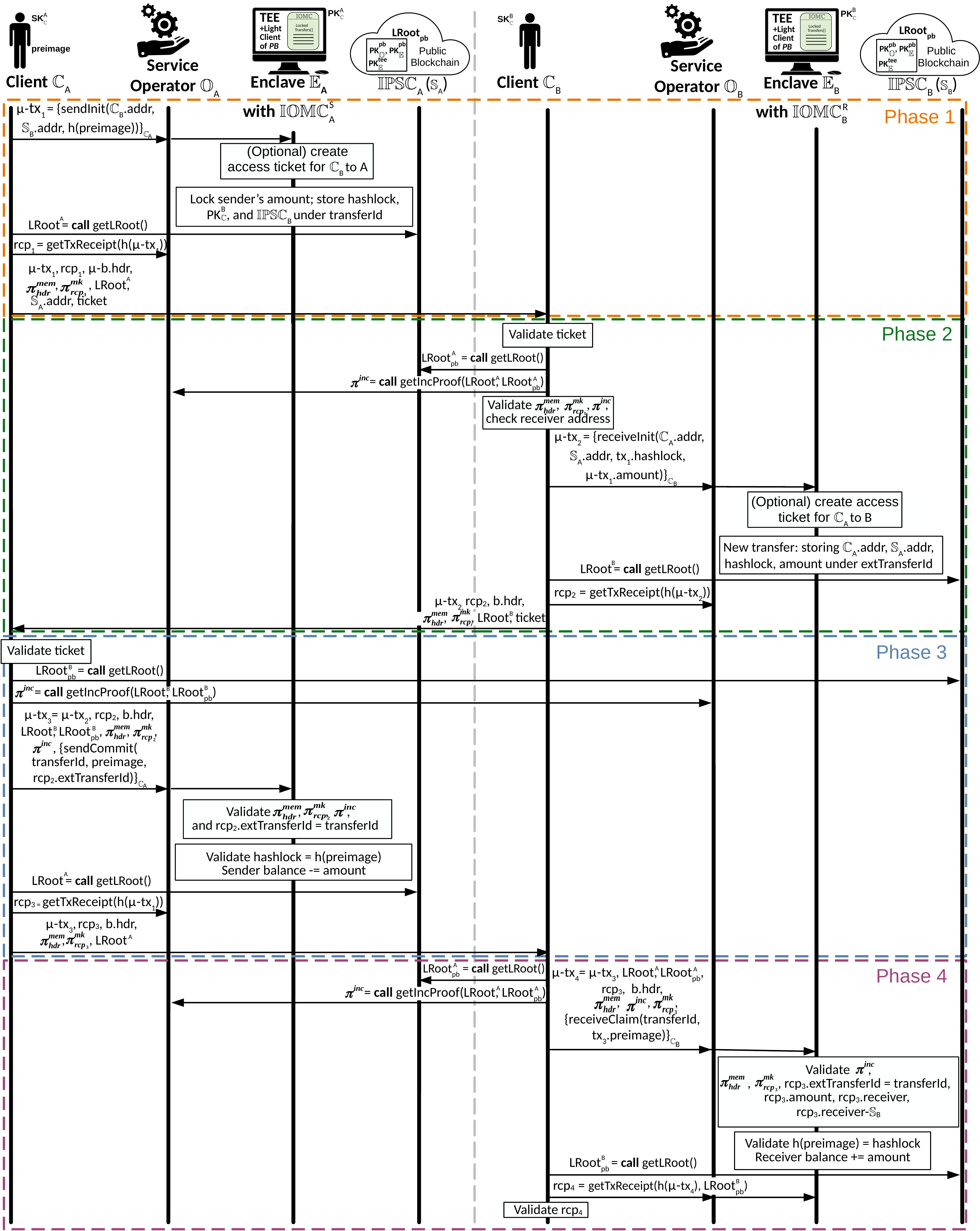}
		\caption{The details of the proposed interoperability protocol $\Pi^{T}$.}
		\label{fig:protocol-full}
	\end{figure*}
	\clearpage

To execute $\Pi^{T}$ it is necessary to inter-connect $\mathbb{E}$s of two instances involved in a transfer.
However, $\mathbb{E}$ does not allow direct communication with the outside world, and therefore it is necessary to use an intermediary.
One solution is to involve $\mathbb{O}$s but they might be overwhelmed with other activities, updating the ledger by executing $\mu$-transactions, and moreover, they might not have direct incentives to execute inter-CBDC transfers.
Therefore, we argue that in contrast to the above option, involving $\mathbb{C}s$ as intermediaries has two advantages: 
(1) elimination of the synchronous communication overhead on $\mathbb{O}$s and 
(2) enabling $\mathbb{C}$s to have a transparent view about the status of the transfer and take action if required.
In the following, we describe phases of $\Pi^T$ in detail (see also \autoref{fig:protocol-full}).

\subsection*{Phase 1 -- Client $\mathbb{C}_A$ Initiates the Protocol}
\label{design:phase1}
The client $\mathbb{C}_A$ creates a $\mu$-tx$_1$ with the amount being sent, which invokes the \texttt{sendInit()} of $\mathbb{IOMC}_A$ with  
arguments containing the address of the external client $\mathbb{C}_B$, the address of $\mathbb{IPSC}_B$ (denoted as $\mathbb{S}_B$ in \autoref{fig:protocol-full} for brevity), and the hash of the secret that is created by $\mathbb{C}_A$.
$\mathbb{C}_A$ sends signed $\mu$-tx$_1$ to $\mathbb{O}_A$ who forwards it to the $\mathbb{E}_A$.
Before executing the $\mu$-tx$_1$, $\mathbb{E}_A$ ensures that the external recipient (i.e., $\mathbb{C}_B$) has the access ticket already issued and valid, enabling her to post censorship resolution requests to $\mathbb{IPSC}_A$ (if needed).
The access ticket should be valid for at least the entire period defined by the HTLC of $\mathbb{IOMC}_A$.
In the next step, a $\mu$-tx$_1$ is executed by $\mathbb{E}_A$, creating a new transfer record with $transferId$ in $\mathbb{IOMC}_A$.
During the execution, $\mathbb{C}_A$'s tokens are transferred (and thus locked) to the $\mathbb{IOMC}_A$'s address.
$\mathbb{C}_A$ waits until the new version of $L_A$ is snapshotted to $\mathbb{IPSC}_A$, and then obtains $LRoot^{A}$ from it.
Then $\mathbb{C}_A$ asks $\mathbb{O}_A$ for the execution receipt $rcp_1$ of $\mu$-tx$_1$ that also contains a set of proofs ($\pi_{hdr}^{mem}$, $\pi_{rcp_1}^{mk}$) and the header of the $\mu$-block that includes  $\mu$-tx$_1$. 
In detail, $\pi_{hdr}^{mem}$ is the inclusion proof of the $\mu$-block \textit{b} in the current version of $L_A$; $\pi_{rcp_1}^{mk}$ is the Merkle proof proving that $rcp_1$ is included in \textit{b} (while $rcp_1$ proves that $\mu$-tx$_1$ was executed correctly).
The mentioned proofs and the receipt are provided to $\mathbb{C}_B$, who verifies that $\mu$-tx$_1$ was executed and included in the $L_A$'s version that is already snapshotted to $\mathbb{IPSC}_A$,  thus irreversible (see below).

\subsection*{Phase 2 -- $\mathbb{C}_B$ Initiates Receive}
\label{design:phase2}
First, $\mathbb{C}_B$ validates an access ticket to $\mathbb{IPSC}_A$ using the enclave $\mathbb{E}_A$'s public key accessible in that smart contract.
Next, $\mathbb{C}_B$ obtains the root hash $LRoot_{pb}^{A}$ of $L_A$ to ensure that $\mathbb{C}_B$'s received state has been already published in $\mathbb{IPSC}_A$, and thus contains $\mu$-tx$_1$.
After obtaining $LRoot_{pb}^{A}$, $\mathbb{C}_B$ forwards it along with the root $LRoot^{A}$ obtained from $\mathbb{C}_A$ to $\mathbb{O}_A$, who creates an incremental proof $\pi^{inc}$ of $\langle LRoot^{A}, LRoot_{pb}^{A}\rangle$.
Once the proof $\pi^{inc}$ has been obtained and validated, the protocol can proceed to validate the remaining proofs sent by the client $\mathbb{C}_A$ along with verifying that the receiving address belongs to $\mathbb{C}_B$.
Next, $\mathbb{C}_B$ creates $\mu$-$tx_2$, invoking the method \texttt{receiveInit()} with the arguments: the address of $\mathbb{C}_A$ obtained from $\mu$-$tx_1$,\footnote{Note that we assume that the address is extractable from the signature.} the address $\mathbb{IPSC}_A.addr$ of $\mathbb{C}_A$'s instance, the hash value of the secret, and the amount of crypto-tokens being sent.
$\mathbb{C}_B$ sends $\mu$-$tx_2$ to $\mathbb{O}_B$, who forwards it to $\mathbb{E}_B$ for processing.
During processing of $\mu$-$tx_2$, $\mathbb{E}_B$ determines whether the external client (from its point of view -- i.e., $\mathbb{C}_A$) has an access ticket issued with a sufficiently long validity period; if not, one is created.
Subsequently, $\mathbb{E}_B$ creates a new record in $\mathbb{IOMC}_B$ with $extTransferId$.
Afterward, $\mathbb{C}_B$ retrieves the $LRoot^{B}$ from $L_B$ and requests the execution receipt  $rcp_2$ from $\mathbb{O}_B$, acknowledging that the $\mu$-$tx_2$ has been executed. 
Finally, $\mathbb{C}_B$ sends a message $\mathbb{C}_A$ with $\mu$-$tx_2$ and cryptographic proofs $\pi_{hdr}^{mem}$, $\pi_{rcp_2}^{mk}$, the execution receipt of $\mu$-$tx_2$, the block header $b$ 
 in which the $\mu$-$tx_2$ was included, $LRoot^B$ (i.e., the root value of $L_B$ after $\mu$-$tx_2$ was executed), and the valid client access ticket for $\mathbb{C}_A$.

\subsection*{Phase 3 -- Confirmation of Transfer by $\mathbb{C}_A$}
\label{design:phase3}
First, $\mathbb{C}_A$ validates the received access ticket to $\mathbb{IPSC}_B$. 
Next, $\mathbb{C}_A$ obtains the snapshotted root hash $LRoot_{pb}^{B}$ of $L_B$ from $\mathbb{IPSC}_B$.
As in the previous phases, it is necessary to verify that the version of $L_B$ that includes $\mu$-$tx_2$ is represented by $LRoot_{pb}^B$ (thus is irreversible). 
Next, both root hashes ($LRoot^B$ and $LRoot_{pb}^B$) are sent to the external operator $\mathbb{O}_B$, which produces the incremental proof $\pi^{inc}$ from them.
Next, $\mathbb{C}_A$ creates $\mu$-$tx_3$ that consists of invoking the \texttt{sendCommit()} method at $\mathbb{E}_A$ with the arguments containing the published secret (i.e., $preimage$) and the record identifier of the transfer at local instance (i.e., $transferId$) as well as the external one (i.e., $extTransferId$).
Along with the invocation of \texttt{sendCommit()}, $\mu$-$tx_3$ also wraps $\pi^{inc}$ with its versions ($LRoot_{pb}^{B}$ and $LRoot^{B}$), $\mu$-$tx_2$, its execution receipt $rcp_2$ with its Merkle proof $\pi_{rcp_2}^{mk}$, $b.hdr$ -- the header of the block that included $\mu$-$tx_2$, and its membership proof $\pi^{mem}_{hdr}$ of $L_B$.
Next, $\mathbb{C}_A$ sends $\mu$-$tx_3$ to $\mathbb{E}_A$ through $\mathbb{O}_A$.
During the execution of $\mu$-$tx_3$, $\mathbb{E}_A$ validates the provided proofs and the equality of transfer IDs from both sides of the protocol.
Note that to verify $\pi^{mem}_{hdr}$, $\mathbb{E}_A$ uses its light client to $L_B$.
$\mathbb{E}_A$ then validates whether $\mathbb{C}_A$'s provided secret corresponds to the hashlock recorded in the 1st phase of the protocol, and if so, it burns the sent balance of the transfer.

Next, $\mathbb{C}_A$ waits until the new version of $L_A$ is snapshotted to $\mathbb{IPSC}_A$, and then obtains $LRoot^{A}$ from it.
Then $\mathbb{C}_A$ asks $\mathbb{O}_A$ for the execution receipt $rcp_3$ of $\mu$-tx$_3$ that also contains a set of proofs ($\pi_{hdr}^{mem}$, $\pi_{rcp_3}^{mk}$) and the header of the $\mu$-block that includes  $\mu$-tx$_3$. 
The proofs have the same interpretation as in the end of the 1st phase. 
The mentioned proofs and the receipt are provided to $\mathbb{C}_B$, who verifies that $\mu$-tx$_1$ was executed and included in the $L_A$'s version that is already snapshotted to $\mathbb{IPSC}_A$,  thus irreversible.

\subsection*{Phase 4 -- Acceptance of Tokens by $\mathbb{C}_B$}
\label{design:phase4}
After receiving a message from client $\mathbb{C}_A$, the client $\mathbb{C}_B$ obtains $LRoot_{pb}^A$ from $\mathbb{IPSC}_A$ and then requests the incremental proof between versions $\langle LRoot^{A}, LRoot_{pb}^{A}\rangle$ from $\mathbb{O}_A$.
Then, $\mathbb{C}_B$ creates $\mu$-$tx_4$ invoking the \texttt{receiveClaim()} function at $\mathbb{E}_B$ with $transferId$ and the disclosed secret by $\mathbb{C}_A$ as the arguments.
Moreover, $\mu$-$tx_4$ contains remaining items received from $\mathbb{C}_A$. 
Then, $\mu$-$tx_4$ is sent to $\mathbb{O}_B$, who forwards it to $\mathbb{E}_B$.
During the execution of $\mu$-$tx_4$, $\mathbb{E}_B$ verifies the provided proofs, the equality of transfer IDs from both sides of the protocol, the amount being sent, and the receiver of the transfer (i.e., $\mathbb{C}_B$ || $\mathbb{IPSC}_B$).
Note that to verify $\pi^{mem}_{hdr}$, $\mathbb{E}_B$ uses its light client to $L_A$.
$\mathbb{E}_A$ then validates whether $\mathbb{C}_A$'s provided secret corresponds to the hashlock recorded in the 2nd phase of the protocol, and if so, it mints the sent balance of the transfer on the receiver's account $\mathbb{C}_B$.
Finally, $\mathbb{C}_B$ verifies that $\mu$-$tx_4$ was snapshotted at $\mathbb{IPSC}_B$, thus is irreversible. 
In detail, first $\mathbb{C}_B$ obtains $LRoot_{pb}^B$ from $\mathbb{IPSC}_B$ and then asks $\mathbb{O}_B$ to provide her with the execution receipt $rcp_4$ of $\mu$-$tx_4$ in the version of $L_B$ that is equal or newer than $LRoot_{pb}^B$. 
Then, $\mathbb{C}_B$ verifies $rcp_4$, which completes the protocol.

 \section{Implementation \& Evaluation}
 \label{sec:eval}
 The work is built on a \textit{proof-of-concept} implementation of the decentralized smart contract platform Aquareum in C++ and Intel SGX technology for enclave instantiation. 
The IPSC contract on the public blockchain is constructed using the Solidity programming language and is prepared for deployment on the Ethereum network.
The enclave employs the OpenEnclave SDK development tool,\footnote{\url{https://openenclave.io/sdk/}}, which is compatible with several TEE technologies and OS systems.
Aquareum incorporates the Ethereum virtual machine -- EVM, in its stripped-down, minimalist version eEVM.\footnote{Microsoft's Enclave EVM is available at https://github.com/microsoft/eEVM.}

\subsubsection{Implementation Details}
The C++ written client application enables the clients to execute internal and external (i.e., between two instance) transfer operations as well as invoking internal and external functions of micro contracts.
The operator component is represented by the C++ written server implemented as a concurrent non-blocking application that processes messages from clients.
So far, the PoC of the server enables to process three types of messages: transaction execution, client registration, query for IOMC contract addresses.

\subsection{Evaluation}
 We used \textit{Ganache}\footnote{\url{https://github.com/trufflesuite/ganache-cli}} and \textit{Truffle},\footnote{\url{https://github.com/trufflesuite/truffle}} to develop $\mathbb{IOMC}$, $\mathbb{IPSC}$, and $\mathbb{IMSC}$ contracts. 
 In addition, using the \textit{Pexpect}\footnote{\url{https://github.com/pexpect/pexpect}} tool, we tested the intercommunication of the implemented components and validated the correctness of the implemented interoperability protocol.
The tool enabled the parallel execution and control of numerous programs (in this case, multiple Aquareum instances and client programs) to check the correctness of the expected output.

The computational cost of executing the operations defined in $\mathbb{IOMC}$ and $\mathbb{IMSC}^X$ contracts is presented in \autoref{table:gas-iomc-send}, \autoref{table:gas-iomc-recv}, and \autoref{table:gas-imsc}.\footnote{Note that we do not provide the gas measurements for $\mathbb{IPSC}$ since these are almost the same as in Aquareum~\cite{homoliak2020aquareum}.}
We optimized our implementation to minimize the storage requirements of smart contract platform.
On the other hand, it is important to highlight that $\mathbb{IOMC}^X$ $\mu$-contracts are executed on a private ledger corresponding to the instance of CBDC, where the cost of gas is minimal or negligible as compared to a public blockchain.
Other experiments  are the subject of our future work.

\begin{table}[t]
\centering
\footnotesize
    \begin{tabular}
    	    {  r  c  c  c  c  } 
    	\toprule
            \textbf{Function} & \vbox{\hbox{{\rotatebox[origin=c]{45}{constructor}}}\vspace{2pt}} & \vbox{\hbox{{\rotatebox[origin=c]{45}{sendInitialize}}}\vspace{2pt}} & \vbox{\hbox{{\rotatebox[origin=c]{45}{sendCommit}}}\vspace{2pt}} & \vbox{\hbox{{\rotatebox[origin=c]{45}{sendRevert}}}\vspace{2pt}}  \\
        \midrule
        	\textbf{Deployment} & 901 509 & 160 698 & 64 629 & 60 923 \\
            \textbf{Execution} & 653 689 & 134 498 & 42 717 & 39 523 \\
        \bottomrule
    \end{tabular}
\caption{The cost of deployment and invocation of functions in the sending $\mathbb{IOMC}^S$ $\mu$-contract in gas units (CBDC private ledger).}
\label{table:gas-iomc-send}
\vspace{-0.4cm}
\end{table}

\begin{table}[t]
    \centering
    \footnotesize
    \renewcommand{\arraystretch}{1.5}
    \begin{tabular}
    {  r  c  c  c  c  } 
        \toprule
            \textbf{Function} & \vbox{\hbox{{\rotatebox[origin]{45}{constructor}}}\vspace{2pt}} & \vbox{\hbox{{\rotatebox[origin]{45}{receiveInit}}}\vspace{2pt}} & \vbox{\hbox{{\rotatebox[origin]{45}{receiveClaim}}}\vspace{2pt}} & \vbox{\hbox{{\rotatebox[origin]{45}{fund}}}\vspace{2pt}}  \\
        \midrule
        	\textbf{Deployment} & 716 330 & 139 218 & 61 245 & 23 168 \\
            \textbf{Execution} & 509 366 & 112 762 & 39 653 & 1 896 \\
        \bottomrule
    \end{tabular}
\caption{The cost of deployment  and invocation of functions in the receiving $\mathbb{IOMC}^R$ $\mu$-contract in units of gas (CBDC private ledger).}
\label{table:gas-iomc-recv}
\vspace{-0.4cm}
\end{table}

\begin{table}[t]
\centering
\footnotesize
    \begin{tabular}
    	    	    {  r  c  c  c  c  } 
    	\toprule
            \textbf{Function} & \vbox{\hbox{{\rotatebox[origin=c]{45}{constructor}}}\vspace{2pt}} & \vbox{\hbox{{\rotatebox[origin=c]{45}{newJoinRequest}}}\vspace{2pt}} & \vbox{\hbox{{\rotatebox[origin=c]{45}{approveRequest}}}\vspace{2pt}} & \vbox{\hbox{{\rotatebox[origin=c]{45}{isApproved}}}\vspace{2pt}}  \\
        \midrule
        	\textbf{Deployment} & 830 074 & 48 629 & 69 642 & 0 \\
        \hline
            \textbf{Execution} & 567 838 & 25 949 & 45 554 & 0 \\
        \bottomrule
    \end{tabular}
\caption{The invocation cost of functions in $\mathbb{IMSC}$ smart contract in units of gas (Ethereum public blockchain).}
\label{table:gas-imsc}
\vspace{-0.4cm}
\end{table}

 \section{Security Analysis}
 \label{sec:secanalysis}
 In this section, we analyze our approach in terms of security-oriented features and requirements specified in \autoref{sec:problem}.
In particular, we focus on resilience analysis of our approach against adversarial actions that the malicious CBDC instance (i.e., its operator $\mathcal{O}$) or malicious client (i.e., $\mathcal{C}$) can perform to violate the security requirements.

\subsection{Single Instance of CBDC}

\begin{theorem}\label{theorem:correctness}
	(Correctness of Operation Execution) 
	$\mathcal{O}$ is unable to modify the full state of $L$ in a way that does not respect the semantics of VM deployed in $\mathbb{E}$ of CBDC instance.
\end{theorem}
\begin{proof1}
	The update of the $L$'s state is performed exclusively in $\mathbb{E}$.	
	Since $\mathbb{E}$ contains trusted code that is publicly known and remotely attested by $\mathbb{C}$s, $\mathcal{O}$ cannot tamper with this code.
\end{proof1}

\begin{theorem}\label{theorem:integrity}
		(Integrity) $\mathcal{O}$ is unable to modify the past records of $L$, and no conflicting transactions can be stored in $L$.
\end{theorem}
\begin{proof1}
	All extensions of $L$ are performed within trusted code of $\mathbb{E}$ (see \autoref{theorem:correctness}), while utilizing the history tree~\cite{crosby2009efficient} as a tamper evident data structure, which enables us to make only such incremental extensions of $L$ that are consistent with $L$'s past. 
\end{proof1}

\begin{theorem}
	(Verifiability) $\mathcal{O}$ is unable to unnoticeably modify or delete a transaction $tx$ that was previously inserted to $L$, if sync with $\mathbb{IPSC}$ was executed anytime afterward.
\end{theorem}

\begin{proof1}
	Since $tx$ was correctly executed (\autoref{theorem:correctness}) as a part of the block $b_i$ in a trusted code of $\mathbb{E}$, $\mathbb{E}$ produced a signed version transition pair $\{h(L_{i-1}), h(L_i), t_i, t_s\}_\mathbb{E}$ of $L$ from the version $i-1$ to the new version $i$ that corresponds to $L$ with $b_i$ included.
	$\mathcal{O}$ could either sync $L$ with $\mathbb{IPSC}$ immediately after $b_i$ was appended or she could do it $n$ versions later.
	In the first case, $\mathcal{O}$ published $\{h(L_{i-1}), h(L_i), t_i, t_s\}_\mathbb{E}$ to $\mathbb{IPSC}$, which updated its current version of $L$ to $i$ by storing $h(L_i)$ into $\mathbb{IPSC}.LRoot_{pb}$. 
	In the second case, $n$ blocks were appended to $L$, obtaining its $(i+n)$th version. 
	$\mathbb{E}$ executed all transactions from versions $(i+1),\ldots, (i+n)$  of $L$, while preserving correctness (\autoref{theorem:correctness}) and integrity (\autoref{theorem:integrity}). 
	Then $\mathbb{E}$ generated a version transition pair $\{h(L_{i-1}), h(L_{i+n}),  t_i, t_s \}_\mathbb{E}$ and $\mathcal{O}$  posted  it to $\mathbb{IPSC}$, where 
	the current version of $L$ was updated to $i+n$ by storing $h(L_{i+n})$ into $\mathbb{IPSC}.LRoot_{pb}$.
	When any $\mathbb{C}$ requests $tx$ and its proofs from $\mathcal{O}$ with regard to publicly visible $\mathbb{IPSC}.LRoot_{pb}$, she might obtain a modified $tx'$ with a valid membership proof $\pi^{mem}_{hdr_i}$ of the block $b_i$ but an invalid Merkle proof  $\pi^{mk}_{tx'}$,  which cannot be forged.~\hfill$\Box$
    \hfill
	In the case of $tx$ deletion, $\mathcal{O}$ provides $\mathbb{C}$ with the tampered full block $b_i'$ (maliciously excluding $tx$) whose membership proof $\pi^{mem}_{hdr_i'}$ is invalid -- it cannot be forged. 	
\end{proof1}

\begin{theorem}\label{theorem:non-equivocation}
	(Non-Equivocation) Assuming $L$ synced with $\mathbb{IPSC}$: $\mathcal{O}$ is unable to provide two distinct $\mathbb{C}$s with two distinct valid views on $L$. 
\end{theorem}
\begin{proof1}
	Since $L$ is periodically synced with publicly accessible $\mathbb{IPSC}$, and $\mathbb{IPSC}$ stores only a single current version of $L$ (i.e., $\mathbb{IPSC}.LRoot_{pb}$), all $\mathbb{C}s$ share the same view on $L$.
\end{proof1}

\begin{theorem}\label{theorem:censorhip}
	(Censorship Evidence) $\mathcal{O}$ is unable to censor any request (transaction or query) from $\mathbb{C}$ while staying unnoticeable.
\end{theorem}
\begin{proof1}
	If $\mathbb{C}$'s request is censored by CBDC's operator $\mathcal{O}$, $\mathbb{C}$ can ask for a resolution of the request through public $\mathbb{IPSC}$.
	$\mathcal{O}$ observing the request might either ignore it and leave the indisputable proof of censorship at $\mathbb{IPSC}$ or she might submit the request to $\mathbb{E}$ and obtain an enclave-signed proof witnessing that a request was processed (hence have not remained censored) -- this proof is submitted to $\mathbb{IPSC}$, whereby publicly resolving the request.
\end{proof1}

\begin{theorem}\label{theorem:privacy}
	(Privacy)  $\mathcal{C}$ is unable to obtain plain text of $\mu$-tran\-sac\-tions of other $\mathbb{C}$s even during the censorship resolution.
\end{theorem}
\begin{proof1}
	$\mu$-transactions are sent to $\mathbb{O}$ in TLS-encrypted messages.
	In the case of censorship resolution, submitted $\mu$-tran\-sac\-tions by $\mathbb{C}$ to public $\mathbb{IPSC}$ are encrypted by $\mathbb{E}$'s public key $PK_{\mathbb{E}}^{tee}$.
\end{proof1}

\begin{theorem}\label{theorem:token-issuance}
	(Transparent Token Issuance)  $\mathcal{O}$ is unable to issue or burn any tokens without leaving a publicly visible evidence.
\end{theorem}
\begin{proof1}
	All issued tokens of CBDC are publicly visible at $\mathbb{IPSC}$ since each transaction posting a new version transition pair also contains $\mathbb{E}$-signed information about the current total issued tokens $t_i$ and total supply of the instance $t_s$,\footnote{Note that in the case of single CBDC instance $t_i = t_s$} while $t_i$ was updated within the trusted code of  $\mathbb{E}$.
	The information about $t_i$ is updated at $\mathbb{IPSC}$ along with the new version of $L$.
	Note that the history of changes in total issued tokens $t_i$ can be parsed from all transactions updating version of $L$ published by $\mathcal{O}$ to $PB$.
\end{proof1}

\subsection{Multiple Instances of CBDC}
In the following, we assume two CBDC instances A and~B.

\begin{theorem}\label{theorem:token-multi-inter}
	(Atomic Interoperability I) Neither $\mathcal{O}_A$ (operating $A$) nor $\mathcal{O}_B$ (operating $B$) is unable to steal any tokens during the inter-bank CBDC transfer. 
\end{theorem}
\begin{proof1}
	Atomic interoperability is ensured in our approach by adaptation of atomic swap protocol for all inter-bank transfers, which enables us to preserve the wholesale environment of CBDC in a consistent state (respecting \autoref{eq:token-sum-multi}).
	In detail, the transferred tokens from CBDC instance $A$ to instance $B$ are not credited to $B$ until $A$ does not provide the indisputable proof that tokens were deducted from a relevant $A$'s account.
	This proof confirms irreversible inclusion of $tx_3$ (i.e., $\mathbb{E}_A.sendCommit()$ that deducts account of $A$'s client) in $A$'s ledger and it is verified in 4th stage of our protocol by the trusted code of $\mathbb{E}_B$.
	
	In the case that $\mathcal{O}_A$ would like to present $B$ with integrity snapshot of $L_A$ that was not synced to  $\mathbb{IPSC}_A$ yet, B will not accept it since the 4th phase of our protocol requires $\mathcal{O}_B$ to fetch the recent $\mathbb{IPSC}_A.LRoot_{pb}$ and verify its consistency with A-provided $LRoot$ as well as inclusion proof in $PB$; all executed/verified within trusted code of $\mathbb{E}_B$.
\end{proof1}

\begin{theorem}\label{theorem:token-multi-inter2}
	(Atomic Interoperability II) Colluding clients $\mathcal{C}_A$ and $\mathcal{C}_B$  of two CBDC instances cannot steal any tokens form the system during the transfer operation of our protocol.
\end{theorem}
\begin{proof1}
    If the first two phases of our protocol have been executed, $\mathcal{C}_A$ might potentially reveal the $preimage$ to $\mathcal{C}_B$ without running the 3rd phase with the intention to credit the tokens at $B$ while deduction at $A$ had not been executed yet.
    However, this is prevented since the trusted code of $\mathbb{E}_B$ verifies that the deduction was performed at $A$ before crediting the tokens to $\mathcal{C}_B$ -- as described in  \autoref{theorem:token-multi-inter}.
\end{proof1}

\begin{theorem}\label{theorem:multi-censorhip1}
	(Inter-CBDC Censorship Evidence) $\mathcal{O}_A$ is unable to unnoticeably censor any request (transaction or query) from $\mathbb{C}_B$.
\end{theorem}
\begin{proof1}
	If $\mathbb{C}_B$'s request is censored by $\mathcal{O}_A$, $\mathbb{C}_B$ can ask for a resolution of the request through public $\mathbb{IPSC}_A$ since $\mathbb{C}_B$ already has the access ticket to instance $A$.
	The access ticket is signed by $\mathbb{E}_A$ and thus can be verified at $\mathbb{IPSC}_A$.
	Hence, the censorship resolution/evidence is the same as in \autoref{theorem:censorhip} of a single CBDC instance.
\end{proof1}

\begin{theorem}\label{theorem:multi-censorship2}
	(Inter-CBDC Censorship Recovery) A permanent inter-CBDC censorship by  $\mathcal{O}_A$ does not cause an inconsistent state or permanently frozen funds of undergoing transfer operations at any other CBDC instance -- all initiated and not finished transfer operations can be recovered from.
	
\end{theorem}
\begin{proof1}
    If $\mathcal{O}_A$ were to censor $\mathbb{C}_B$ in the 2nd phase of our protocol, no changes at ledger $L_B$ would be made.
    If $\mathcal{O}_A$ were to censor $\mathbb{C}_B$ in the 4th phase of our protocol, $L_B$ would contain an initiated transfer entry, which has not any impact on the consistency of the ledger since it does not contain any locked tokens.~\hfill$\Box$

    \smallskip \noindent
    If $\mathcal{O}_B$ were to censor $\mathbb{C}_A$ in the 3rd phase of our protocol, $A$ would contain some frozen funds of the initiated transfer.
    However, these funds can be recovered back to $\mathbb{C}_A$ upon a recovery call of $\mathbb{E}_A$ after a recovery timeout has passed.
    Note that after tokens of $\mathbb{C}_A$ have been recovered and synced to $\mathbb{IPSC}_A$ in $PB$, it is not possible to finish the 4th stage of our protocol since it requires providing the proof that tokens were deducted at $A$ and such a proof cannot be constructed anymore.
    The same holds in the situation where the sync to  $\mathbb{IPSC}_A$ at $PB$ has not been made yet -- after recovery of tokens, $\mathbb{E}_A$ does not allow to deduct the same tokens due to its correct execution (see \autoref{theorem:correctness}). 
\end{proof1}

\begin{theorem}\label{theorem:multi-identity-1}
    (Identity Management of CBDC Instances I) A new (potentially fake) CBDC instance cannot enter the ecosystem of wholesale CBDC upon its decision.
\end{theorem}
\begin{proof1}
    To extend the list of valid CBDC instances (stored in IMSC contract), the majority vote of all existing CBDC instances must be achieved through public voting on IMSC.
\end{proof1}

\begin{theorem}\label{theorem:multi-identity-2}
    (Identity Management of CBDC Instances II) Any CBDC instance (that e.g., does not respect certain rules for issuance of tokens) might be removed from the ecosystem of CBDC by majority vote.
\end{theorem}
\begin{proof1}
    A publicly visible voting about removal of a CBDC instance from the ecosystem is realized by IMSC contract that resides in $PB$, while each existing instance has a single vote. 
\end{proof1}

\subsection{\textbf{Security of TEE}}
We assume that its TEE platform employed is secure.
However, previous research indicated that this might not be the case in practical implementations of TEE, such as SGX that was vulnerable to memory corruption attacks~\cite{biondo2018guard} as well as side channel attacks~\cite{brasser2017dr,van2018foreshadow,Lipp2021Platypus,Murdock2019plundervolt}.
A number of software-based defense and mitigation techniques have been  proposed~\cite{shih2017t,gruss2017strong,chen2017detecting,brasser2017dr,seo2017sgx} and some vulnerabilities were patched by Intel at the hardware level~\cite{intel-sgx-response}.
Nevertheless, we note that our approach is TEE-agnostic thus can be integrated with other TEEs such as ARM TrustZone or RISC-V architectures (using Keystone-enclave~\cite{Keystone-enclave} or Sanctum~\cite{costan2016sanctum}).

Another class of SGX vulnerabilities was presented by Cloosters et al.~\cite{cloosters2020teerex} and involved incorrect application designs enabling arbitrary reads and writes of protected memory.
Since the authors did not provide public with their tool (and moreover it does not support Open-enclave SDK), we did manual inspection of our code and did not find any of the concerned vulnerabilities. 
Another work was done by Borrello et al.~\cite{Borrello2022AEPIC} and involves more serious micro-architectural flaws in chip design.  
Intel has already released microcode and SGX SDK updates to fix the issue.

\subsection{\textbf{Public Blockchain \& Finality}}
Many blockchain platforms suffer from accidental forks (i.e., availa\-bi\-lity-favored blockchains in terms of CAP theorem), which temporarily create parallel inconsistent blockchain views.
To mitigate this phenomenon, it is recommended to wait a certain number of block confirmations after a given block is created before considering it irreversible with overwhelming probability.
This waiting time (a.k.a., time to finality) influences the non-equivocation property of our approach, inheriting it from the underlying blockchain platform.
Most availability-favored blockchains have a long time to finality, e.g., $\sim$3mins in Bitcoin~\cite{nakamoto2008bitcoin}, $\sim$3mins in Ethereum~\cite{wood2014ethereum}, $\sim$2mins in Cardano~\cite{kiayias2017ouroboros}.
However, consistency-favored blockchains in terms of the CAP theorem have a short time to finality, e.g., HoneyBadgerBFT~\cite{miller2016honey}, Algorand~\cite{gilad2017algorand}, Hyperledger Besu~\cite{hyperledger-github}. 
The selection of the underlying blockchain platform should respect low time to finality in the critical environment of CBDC, and thus employ a consistency-favored public blockchain.

 \section{Related Work}
 \label{sec:related}
 In this section, we first review various approaches to interoperability and CBDC. Moreover, since our protocol is designed using a combination of TEE and the blockchain, we revise the most relevant solutions and stress the novelty of our approach, which combines several unique features.

\subsection{Blockchain Interoperability}

Cross-chain interoperability is one of the most desirable yet challenging features to be designed and developed in blockchains, affecting the impact and usability of the solution~\cite{wang2023exploring,belchior2021survey,mohanty2022blockchain,qasse2019inter}.

Cross-chain communication protocols define the process of synchronization between different chains of the same blockchain, e.g., by the use of sidechains.
Additionally, cross-blockchain communication protocols, such as Interledger Protocol~\cite{2019Interledger}, allow interaction of different blockchains. 
While the cross-chain solutions can be employed by the native constructs such as atomic swap, the cross-blockchain protocols require adoption of the solution.
Blockchain interoperability solutions can be categorized into three groups according to the principle they are based on and the type of chains that are supported~\cite{belchior2021survey}.

\paragraph{\textbf{Public connectors}.} Public connectors are a set of approaches that focuses on cryptocurrency systems and their transactions. This includes the sidechains, relays, notary schemes, and hash time locks~\cite{2019SideChains,mohanty2022blockchain,2021HashTimeLocks}.

\paragraph{\textbf{Blockchain of blockchains}.} Blockchain of blockchains focuses on application specific-solutions. The example is Polkadot~\cite{belchior2021survey, 2020Polkadot}~--~a network for cross-blockchain interoperability. In Polkadot network, multiple parallelized globally-coherent chains (parachains) are connected via bridges that represent a specific type of parachain. 
Bridges also serve as a gateway for communication with external networks, such as Bitcoin. 

\paragraph{\textbf{Hybrid Connectors}} Hybrid solutions create an abstraction layer over the blockchain ecosystem and provide a unified API for interaction between blockchain and applications~\cite{2021aapXChain}. Examples are trusted relays or blockchain migrators. The interoperability requires validators present in both the source and target blockchains. The validators collect cross-chain transactions and ensure that they are delivered \cite{2021TrustedRelays}.\\

The proposed solution in this paper contain a custom one-way atomic swap protocol that utilizes hash time lock contracts. 
Such swaps are settled on public blockchain $PB$. 
It is also expected that $PB$ used for the synchronization of clients and CBDC instances deploys a single blockchain technology. The usability of the proposed solution targets the financial institutions such as banks, leveraging its potential use in CBDC projects. 
The protocol does not specify a middleware layer providing API or the use of gateway chains. 
Therefore, it can be categorized as a public connector that augments and combines the features provided by individual solutions in the same category.

\subsection{\textbf{CBDC Projects}}
\label{sec:cbdcprojects}
While most CBDC projects are still in their early stages, some well-known proposals are reaching maturity level~\cite{zhang2021blockchain}. For instance, Project Jasper \cite{chapman2017project} was one of the initial prototypes for inter-bank payments using blockchain technology. 
Project Ubin \cite{ubin} appeared with the aim of clearing and settling of payments and securities efficiently by using several blockchain technologies and smart contracts. 
Project E-krona~\cite{armelius2020krona} was designed, among others, to enable fast transactions between domestic and cross-border entities. 
Stella~\cite{kishi2019project} is another well-known project that uses permissioned blockchain technology to enable cross border operations as well as confidentiality protection. 
The mBridge project~\cite{mbridge} (initially named Inthanon-LionRock) prototype is built by ConsenSys on Hyperledger Besu. The prototype encompasses several jurisdictions and aims at creating a cross-border payment infrastructure that improves on key pain points, including high cost, low speed, and operational complexities. Finally, Project Khokha~\cite{Khokha} was designed for efficient, confidential inter-bank transactions.

\medskip
Despite the maturity of some projects, research on CBDC technology is still in its infancy. 
In addition, the road to creating a native interoperable protocol that can be used regardless of the underlying blockchain technology still requires further exploration and is one of the main objectives of this article. 
Compared to other CBDC projects, our approach is the first protocol combining TEE and blockchain to bring interesting security and privacy features, accompanied by external interoperability. 
In detail, our protocol guarantees a set of features such as integrity, non-equivocation (i.e., we provide snapshots to public blockchain to avoid reverts and forks of the local CBDC ledgers), correctness (i.e., the EVM is executed in an enclave which can be remotely attested), and censorship evidence.
Since the designed protocol addresses inter-bank communication and payment settlements, it can be potentially integrated as a part of the above-mentioned wholesale CBDC projects. The advantages of the retail CBDC approach towards individual clients are also preserved by the privacy support, censorship evidence and mitigation of malicious approach described in Section \ref{sec:secanalysis}. The general approach is also invariant towards token differences introduced by different projects with regard to the public blockchain.

\subsection{\textbf {Combining Blockchain and TEE}}

The combination of Trusted Execution Environment (TEE) technologies and blockchain has gained increased attention in the past few years. Hybridchain \cite{wang2020hybridchain} is an architecture for confidentiality-preserving in permissioned blockchain. Such architecture extends the enclave memory of TEE that allows blockchain applications running in TEE to securely store transaction records outside of TEE. Ekiden \cite{cheng2018ekiden} is a blockchain-agnostic solution that offloads smart contract execution to TEE enclaves.  
Teechain \cite{lind2017teechain} focuses on the Bitcoin network and enables the secure execution of transactions in TEE, enhancing the scalability of the network. Fastkitten enables extended functionality in the Bitcoin network by using Turing-complete smart contracts executed via TEE-enabled operators~\cite{das2019fastkitten}. 

However, solutions combining interoperability with TEE-based blockchains are still in their infancy. Only a few authors have explored this such as Bellavista et al.~\cite{bellavista2021interoperable}, and Lan et al.~\cite{lan2021trustcross}, which are the works most similar to ours. 
More concretely, Bellavista et al.~\cite{bellavista2021interoperable} explore the use of a relay scheme based on TEE to provide blockchain interoperability in the context of collaborative manufacturing and supply chains. 
Lan et al.~\cite{lan2021trustcross} aim to preserve confidentiality in interoperable cross-chain platforms and propose a protocol to ensure privacy-preserving communications among them. 
Nevertheless, our approach is the first one designing a functional protocol for interoperable CBDC, considering features such as the ones mentioned in \autoref{sec:problem}.

 \section{Discussion}
 \label{sec:discussion}

As seen in Section \ref{sec:related}, this is the first blockchain TEE-based interoperable protocol that operates in the context of CBDC. However, our protocol allows modifications if additional requirements were to be fulfilled (i.e., considering the ones defined in Section \ref{sec:problem}). The latter enables a certain degree of dynamism when adapting the protocol to specific application contexts. 

Following the interest of countries in CBDC \cite{cbdctracker}, research on CBDCs and their potential challenges has also been receiving increasing attention in the last years (i.e., the number of contributions has been doubling yearly since 2020 according to Scopus, using the query TITLE-ABS-KEY ( ( ( central  AND  bank  AND  digital  AND  currency )  OR  CBDC )  AND  challenges ). While a profound analysis of state of the art is out of the scope of this paper, we found that authors typically follow two strategies to discuss CBDC and its challenges, namely considering a local perspective (i.e., at a jurisdiction or national level) and adopting a global challenge abstraction. Overall, we considered the most recent reviews and surveys analyzing CBDC and its challenges~\cite{Catalini2022117,Alwago2022553,Koziuk202112,Sebastiao2021305} and other grey literature, such as the Digital Euro Association~\cite{deass_cbdc}, or the US federal reserve \cite{federalreserve}, to extract the challenges and represent them according to a high-level hierarchical abstraction. Since one of the aims of our proposal is to provide solutions to as many challenges as possible, we describe, for each challenge, the benefits and features that our proposal provides in Table \ref{tab:challenges}.

\begin{table*}[t]
	\footnotesize
\begin{tabular}{p{0.9in}  p{2.5in} p{3in} }
\toprule
\textbf{Topic}              & \textbf{Main concerns}                                                                                                                                                                              & \textbf{Our proposal's contribution} \\ 
\midrule
Technology                  & The design, implementation and maintenance of CBDC's as well as their scalability, resiliency and compatibility with the current financial structure.                                               &       Our system is scalable and compatible with current financial system                               \\ \hline
Monetary Policy             & Monetary policy transmission, including interest rates, the value of money, or other tools, should not be hindered by CBDC.                                                                         &      Our system relies on smart contracts to enforce specific policies if required, such as token expiration or token usability.                                \\ \hline
Financial Stability         & The potentially disrupting impact of CBDC on the existing financial system could create new financial vulnerabilities, uncontrolled disintermediation or illicit activities.                 &         The use of blockchain and the policies translated into the system should be audited and verified. Our system is compatible with the latter and other policies in the above layers.                              \\ \hline
Legal Framework             & The legal framework for CBDC needs to comply with existing laws and regulations, including consumer protection, anti-money laundering, and countering the financing of terrorism. &   The system is compatible with auditability layers compliant with current legal and regulations                                    \\ \hline
Interoperability            & Ensuring interoperability by guaranteeing that CBDCs are compatible with other countries monetary policies and promoting cross-border cooperation and standardisation.                             &       Our protocol is interoperable by design and ensures the system remains in monetary equilibrium since no new tokens are created. The potential use of oracles enables further operations with different currencies beyond current ones, promoting cross-border cooperation and additional capabilities.                                \\ \hline
Security and Privacy       & CBDC needs to ensure robustness to prevent cyberattacks and unauthorised access to data by guaranteeing privacy-preserving mechanisms of transactions and personal information.                     &       Our proposal is robust and preserves the privacy of transactions since all the transactions are encrypted. We provide various security properties, such as atomicity, verifiability, integrity, non-equivocation, correctness of execution, censorship evidence, and others.                      \\ \hline
User Adoption and Inclusion & CBDCs will need to provide access to banking services to different populations. Users will require a behaviour change, acceptance and trust.                                               &    The use of our system is transparent to other layers, so it does not introduce any burden. TEE technologies enable trustable platforms, and our protocol allows verifiable censorship resolution.                              \\ 
\bottomrule

\end{tabular}
\caption{High-level abstraction of CBDC's challenges and how our proposal contributes to them. In some cases, our proposal slightly interferes with these challenges since many only apply to other CBDC ecosystem layers.}
\label{tab:challenges}

\end{table*}

 \section{Conclusion}
 \label{sec:conclusion}

Although the controversy surrounding the coexistence of privacy and CBDC \cite{cbdc_privacy_disc}, the latter promises a series of benefits, such as transaction efficiency (e.g., by reducing costs and decreasing its finality at the national or international level) and countering financial crime. Moreover, CBDC complements current financial services by offering broader opportunities. Nevertheless, 
the corresponding regulations should carefully manage these new opportunities, ensuring they do not restrict citizens' rights. Note that novel functionalities enforced in financial transactions, such as token expiration dates, negative interest rates for token holders (i.e., in an attempt to stimulate the economy in recession periods) or tokens whose validity is tied to a specific subset of goods (e.g., enforcing that part of the salary is spent on energy or healthcare), could either be applied for the sustainability of the society or state control in the context of authoritarian regimes. 

Given the above circumstances, we provide the design and implementation of the protocol that uses a custom adaptation of atomic swap and is executed by any pair of CBDC instances to realize a one-way transfer, resolving interoperability over multiple instances of semi-centralized CBDC. Our protocol guarantees a series of properties such as verifiability, atomicity of inter-bank transfers, censorship resistance, and privacy. Our contributions result in a step forward toward enriching the capabilities of CBDC and their practical deployment. 

Future work will closer study token issuance management through protocol directives, perform more extensive evaluation, and propose interoperable execution of smart contracts between CBDC instances.

 \section*{Acknowledgments}
This work was supported by the FIT BUT internal project FIT-S-23-8151.
This work was also supported by the European Commission under the Horizon Europe Programme, as part of the LAZARUS project (\url{http://lazarus-he.eu/} Grant Agreement no. 101070303). Fran Casino was supported by the Government of Catalonia with the Beatriu de Pinós programme (Grant No. 2020 BP 00035), and by AGAUR with the 2021-SGR-00111 project. 
The content of this article does not reflect the official opinion of the European Union. Responsibility for the information and views expressed therein lies entirely with the authors.

\bibliographystyle{ACM-Reference-Format}
\bibliography{refs}


\begin{thebibliography}{71}


\ifx \showCODEN    \undefined \def \showCODEN     #1{\unskip}     \fi
\ifx \showDOI      \undefined \def \showDOI       #1{#1}\fi
\ifx \showISBNx    \undefined \def \showISBNx     #1{\unskip}     \fi
\ifx \showISBNxiii \undefined \def \showISBNxiii  #1{\unskip}     \fi
\ifx \showISSN     \undefined \def \showISSN      #1{\unskip}     \fi
\ifx \showLCCN     \undefined \def \showLCCN      #1{\unskip}     \fi
\ifx \shownote     \undefined \def \shownote      #1{#1}          \fi
\ifx \showarticletitle \undefined \def \showarticletitle #1{#1}   \fi
\ifx \showURL      \undefined \def \showURL       {\relax}        \fi
\providecommand\bibfield[2]{#2}
\providecommand\bibinfo[2]{#2}
\providecommand\natexlab[1]{#1}
\providecommand\showeprint[2][]{arXiv:#2}

\bibitem[\protect\citeauthoryear{AlShamsi, Al-Emran, and Shaalan}{AlShamsi
  et~al\mbox{.}}{2022}]%
        {alshamsi2022systematic}
\bibfield{author}{\bibinfo{person}{Mohammed AlShamsi}, \bibinfo{person}{Mostafa
  Al-Emran}, {and} \bibinfo{person}{Khaled Shaalan}.}
  \bibinfo{year}{2022}\natexlab{}.
\newblock \showarticletitle{A Systematic Review on Blockchain Adoption}.
\newblock \bibinfo{journal}{\emph{Applied Sciences}} \bibinfo{volume}{12},
  \bibinfo{number}{9} (\bibinfo{year}{2022}), \bibinfo{pages}{4245}.
\newblock


\bibitem[\protect\citeauthoryear{Alwago}{Alwago}{2022}]%
        {Alwago2022553}
\bibfield{author}{\bibinfo{person}{W.O. Alwago}.}
  \bibinfo{year}{2022}\natexlab{}.
\newblock \showarticletitle{Is the Renminbi a Global Currency in the Making?
  Globalization of Digital yuan}.
\newblock  \bibinfo{volume}{67}, \bibinfo{number}{4} (\bibinfo{year}{2022}),
  \bibinfo{pages}{553--566}.
\newblock


\bibitem[\protect\citeauthoryear{Anati, Gueron, Johnson, and Scarlata}{Anati
  et~al\mbox{.}}{2013}]%
        {anati2013innovative}
\bibfield{author}{\bibinfo{person}{Ittai Anati}, \bibinfo{person}{Shay Gueron},
  \bibinfo{person}{Simon Johnson}, {and} \bibinfo{person}{Vincent Scarlata}.}
  \bibinfo{year}{2013}\natexlab{}.
\newblock \showarticletitle{Innovative technology for CPU based attestation and
  sealing}. In \bibinfo{booktitle}{\emph{Proceedings of the 2nd international
  workshop on hardware and architectural support for security and privacy}},
  Vol.~\bibinfo{volume}{13}. ACM New York, NY, USA.
\newblock


\bibitem[\protect\citeauthoryear{Armelius, Guibourg, Johansson, and
  Schmalholz}{Armelius et~al\mbox{.}}{2020}]%
        {armelius2020krona}
\bibfield{author}{\bibinfo{person}{Hanna Armelius}, \bibinfo{person}{Gabriela
  Guibourg}, \bibinfo{person}{Stig Johansson}, {and} \bibinfo{person}{Johan
  Schmalholz}.} \bibinfo{year}{2020}\natexlab{}.
\newblock \showarticletitle{E-krona design models: pros, cons and trade-offs}.
\newblock \bibinfo{journal}{\emph{Sveriges Riksbank Economic Review}}
  \bibinfo{volume}{2} (\bibinfo{year}{2020}), \bibinfo{pages}{80--96}.
\newblock


\bibitem[\protect\citeauthoryear{{Atlantic Council}}{{Atlantic
  Council}}{[n.\,d.]}]%
        {cbdctracker}
\bibfield{author}{\bibinfo{person}{{Atlantic Council}}.}
  \bibinfo{year}{[n.\,d.]}\natexlab{}.
\newblock \bibinfo{title}{{Central Bank Digital Currency Tracker}}.
\newblock
  \bibinfo{howpublished}{\url{https://www.atlanticcouncil.org/cbdctracker/}}.
\newblock


\bibitem[\protect\citeauthoryear{Belchior, Vasconcelos, Guerreiro, and
  Correia}{Belchior et~al\mbox{.}}{2021}]%
        {belchior2021survey}
\bibfield{author}{\bibinfo{person}{Rafael Belchior}, \bibinfo{person}{Andr{\'e}
  Vasconcelos}, \bibinfo{person}{S{\'e}rgio Guerreiro}, {and}
  \bibinfo{person}{Miguel Correia}.} \bibinfo{year}{2021}\natexlab{}.
\newblock \showarticletitle{A survey on blockchain interoperability: Past,
  present, and future trends}.
\newblock \bibinfo{journal}{\emph{ACM Computing Surveys (CSUR)}}
  \bibinfo{volume}{54}, \bibinfo{number}{8} (\bibinfo{year}{2021}),
  \bibinfo{pages}{1--41}.
\newblock


\bibitem[\protect\citeauthoryear{Bellavista, Esposito, Foschini, Giannelli,
  Mazzocca, and Montanari}{Bellavista et~al\mbox{.}}{2021}]%
        {bellavista2021interoperable}
\bibfield{author}{\bibinfo{person}{Paolo Bellavista},
  \bibinfo{person}{Christian Esposito}, \bibinfo{person}{Luca Foschini},
  \bibinfo{person}{Carlo Giannelli}, \bibinfo{person}{Nicola Mazzocca}, {and}
  \bibinfo{person}{Rebecca Montanari}.} \bibinfo{year}{2021}\natexlab{}.
\newblock \showarticletitle{Interoperable Blockchains for Highly-Integrated
  Supply Chains in Collaborative Manufacturing}.
\newblock \bibinfo{journal}{\emph{Sensors}} \bibinfo{volume}{21},
  \bibinfo{number}{15} (\bibinfo{year}{2021}), \bibinfo{pages}{4955}.
\newblock


\bibitem[\protect\citeauthoryear{Biondo, Conti, Davi, Frassetto, and
  Sadeghi}{Biondo et~al\mbox{.}}{2018}]%
        {biondo2018guard}
\bibfield{author}{\bibinfo{person}{Andrea Biondo}, \bibinfo{person}{Mauro
  Conti}, \bibinfo{person}{Lucas Davi}, \bibinfo{person}{Tommaso Frassetto},
  {and} \bibinfo{person}{Ahmad-Reza Sadeghi}.} \bibinfo{year}{2018}\natexlab{}.
\newblock \showarticletitle{The Guard's Dilemma: Efficient Code-Reuse Attacks
  Against Intel $\{$SGX$\}$}. In \bibinfo{booktitle}{\emph{27th $\{$USENIX$\}$
  Security Symposium ($\{$USENIX$\}$ Security 18)}}.
  \bibinfo{pages}{1213--1227}.
\newblock


\bibitem[\protect\citeauthoryear{{BIS innovation hub}}{{BIS innovation
  hub}}{2021}]%
        {mbridge}
\bibfield{author}{\bibinfo{person}{{BIS innovation hub}}.}
  \bibinfo{year}{2021}\natexlab{}.
\newblock \bibinfo{title}{{Inthanon-LionRock to mBridge: Building a multi CBDC
  platform for international payments}}.
\newblock
\newblock
\urldef\tempurl%
\url{https://www.bis.org/publ/othp40.htm}
\showURL{%
\tempurl}


\bibitem[\protect\citeauthoryear{{Bitcoin Wiki}}{{Bitcoin Wiki}}{2018}]%
        {atomic-swap}
\bibfield{author}{\bibinfo{person}{{Bitcoin Wiki}}.}
  \bibinfo{year}{2018}\natexlab{}.
\newblock \bibinfo{title}{{Atomic Swap}}.
\newblock
\newblock
\urldef\tempurl%
\url{{https://en.bitcoinwiki.org/wiki/Atomic_Swap}}
\showURL{%
\tempurl}


\bibitem[\protect\citeauthoryear{Boar and Wehrli}{Boar and Wehrli}{2021}]%
        {boar2021ready}
\bibfield{author}{\bibinfo{person}{Codruta Boar} {and} \bibinfo{person}{Andreas
  Wehrli}.} \bibinfo{year}{2021}\natexlab{}.
\newblock \showarticletitle{Ready, steady, go?-Results of the third BIS survey
  on central bank digital currency}.
\newblock  (\bibinfo{year}{2021}).
\newblock


\bibitem[\protect\citeauthoryear{Borrello, Kogler, Schwarzl, Lipp, Gruss, and
  Schwarz}{Borrello et~al\mbox{.}}{2022}]%
        {Borrello2022AEPIC}
\bibfield{author}{\bibinfo{person}{Pietro Borrello}, \bibinfo{person}{Andreas
  Kogler}, \bibinfo{person}{Martin Schwarzl}, \bibinfo{person}{Moritz Lipp},
  \bibinfo{person}{Daniel Gruss}, {and} \bibinfo{person}{Michael Schwarz}.}
  \bibinfo{year}{2022}\natexlab{}.
\newblock \showarticletitle{{ÆPIC Leak}: Architecturally Leaking Uninitialized
  Data from the Microarchitecture}. In \bibinfo{booktitle}{\emph{31st USENIX
  Security Symposium (USENIX Security 22)}}.
\newblock


\bibitem[\protect\citeauthoryear{Brasser, Capkun, Dmitrienko, Frassetto,
  Kostiainen, M{\"u}ller, and Sadeghi}{Brasser et~al\mbox{.}}{2017}]%
        {brasser2017dr}
\bibfield{author}{\bibinfo{person}{Ferdinand Brasser}, \bibinfo{person}{Srdjan
  Capkun}, \bibinfo{person}{Alexandra Dmitrienko}, \bibinfo{person}{Tommaso
  Frassetto}, \bibinfo{person}{Kari Kostiainen}, \bibinfo{person}{Urs
  M{\"u}ller}, {and} \bibinfo{person}{Ahmad-Reza Sadeghi}.}
  \bibinfo{year}{2017}\natexlab{}.
\newblock \showarticletitle{DR. SGX: hardening SGX enclaves against cache
  attacks with data location randomization}.
\newblock \bibinfo{journal}{\emph{arXiv preprint arXiv:1709.09917}}
  (\bibinfo{year}{2017}).
\newblock


\bibitem[\protect\citeauthoryear{Brown, Carlyle, Grigg, and Hearn}{Brown
  et~al\mbox{.}}{2016}]%
        {brown2016corda}
\bibfield{author}{\bibinfo{person}{Richard~Gendal Brown},
  \bibinfo{person}{James Carlyle}, \bibinfo{person}{Ian Grigg}, {and}
  \bibinfo{person}{Mike Hearn}.} \bibinfo{year}{2016}\natexlab{}.
\newblock \showarticletitle{Corda: an introduction}.
\newblock \bibinfo{journal}{\emph{R3 CEV, August}} \bibinfo{volume}{1},
  \bibinfo{number}{15} (\bibinfo{year}{2016}), \bibinfo{pages}{14}.
\newblock


\bibitem[\protect\citeauthoryear{Burdges, Cevallos, Czaban, Habermeier,
  Hosseini, Lama, Alper, Luo, Shirazi, Stewart, and Wood}{Burdges
  et~al\mbox{.}}{2020}]%
        {2020Polkadot}
\bibfield{author}{\bibinfo{person}{Jeff Burdges}, \bibinfo{person}{Alfonso
  Cevallos}, \bibinfo{person}{Peter Czaban}, \bibinfo{person}{Rob Habermeier},
  \bibinfo{person}{Syed Hosseini}, \bibinfo{person}{Fabio Lama},
  \bibinfo{person}{Handan~Kilinc Alper}, \bibinfo{person}{Ximin Luo},
  \bibinfo{person}{Fatemeh Shirazi}, \bibinfo{person}{Alistair Stewart}, {and}
  \bibinfo{person}{Gavin Wood}.} \bibinfo{year}{2020}\natexlab{}.
\newblock \bibinfo{title}{Overview of Polkadot and its Design Considerations}.
\newblock
\newblock
\urldef\tempurl%
\url{https://doi.org/10.48550/ARXIV.2005.13456}
\showDOI{\tempurl}


\bibitem[\protect\citeauthoryear{Catalini, De~Gortari, and Shah}{Catalini
  et~al\mbox{.}}{2022}]%
        {Catalini2022117}
\bibfield{author}{\bibinfo{person}{C. Catalini}, \bibinfo{person}{A.
  De~Gortari}, {and} \bibinfo{person}{N. Shah}.}
  \bibinfo{year}{2022}\natexlab{}.
\newblock \showarticletitle{Some Simple Economics of Stablecoins}.
\newblock   \bibinfo{volume}{14} (\bibinfo{year}{2022}),
  \bibinfo{pages}{117--135}.
\newblock


\bibitem[\protect\citeauthoryear{Chapman, Garratt, Hendry, McCormack, and
  McMahon}{Chapman et~al\mbox{.}}{2017}]%
        {chapman2017project}
\bibfield{author}{\bibinfo{person}{James Chapman}, \bibinfo{person}{Rodney
  Garratt}, \bibinfo{person}{Scott Hendry}, \bibinfo{person}{Andrew McCormack},
  {and} \bibinfo{person}{Wade McMahon}.} \bibinfo{year}{2017}\natexlab{}.
\newblock \showarticletitle{Project Jasper: Are distributed wholesale payment
  systems feasible yet}.
\newblock \bibinfo{journal}{\emph{Financial System}}  \bibinfo{volume}{59}
  (\bibinfo{year}{2017}).
\newblock


\bibitem[\protect\citeauthoryear{Chen, Zhang, Reiter, and Zhang}{Chen
  et~al\mbox{.}}{2017}]%
        {chen2017detecting}
\bibfield{author}{\bibinfo{person}{Sanchuan Chen}, \bibinfo{person}{Xiaokuan
  Zhang}, \bibinfo{person}{Michael~K Reiter}, {and} \bibinfo{person}{Yinqian
  Zhang}.} \bibinfo{year}{2017}\natexlab{}.
\newblock \showarticletitle{Detecting privileged side-channel attacks in
  shielded execution with D{\'e}j{\'a} Vu}. In
  \bibinfo{booktitle}{\emph{Proceedings of the 2017 ACM on Asia Conference on
  Computer and Communications Security}}. \bibinfo{pages}{7--18}.
\newblock


\bibitem[\protect\citeauthoryear{Cheng, Zhang, Kos, He, Hynes, Johnson, Juels,
  Miller, and Song}{Cheng et~al\mbox{.}}{2018}]%
        {cheng2018ekiden}
\bibfield{author}{\bibinfo{person}{Raymond Cheng}, \bibinfo{person}{Fan Zhang},
  \bibinfo{person}{Jernej Kos}, \bibinfo{person}{Warren He},
  \bibinfo{person}{Nicholas Hynes}, \bibinfo{person}{Noah Johnson},
  \bibinfo{person}{Ari Juels}, \bibinfo{person}{Andrew Miller}, {and}
  \bibinfo{person}{Dawn Song}.} \bibinfo{year}{2018}\natexlab{}.
\newblock \showarticletitle{Ekiden: A platform for confidentiality-preserving,
  trustworthy, and performant smart contract execution}.
\newblock \bibinfo{journal}{\emph{arXiv preprint arXiv:1804.05141}}
  (\bibinfo{year}{2018}).
\newblock


\bibitem[\protect\citeauthoryear{Cloosters, Rodler, and Davi}{Cloosters
  et~al\mbox{.}}{2020}]%
        {cloosters2020teerex}
\bibfield{author}{\bibinfo{person}{Tobias Cloosters}, \bibinfo{person}{Michael
  Rodler}, {and} \bibinfo{person}{Lucas Davi}.}
  \bibinfo{year}{2020}\natexlab{}.
\newblock \showarticletitle{TeeRex: Discovery and Exploitation of Memory
  Corruption Vulnerabilities in $\{$SGX$\}$ Enclaves}. In
  \bibinfo{booktitle}{\emph{29th $\{$USENIX$\}$ Security Symposium
  ($\{$USENIX$\}$ Security 20)}}. \bibinfo{pages}{841--858}.
\newblock


\bibitem[\protect\citeauthoryear{Costan, Lebedev, and Devadas}{Costan
  et~al\mbox{.}}{2016}]%
        {costan2016sanctum}
\bibfield{author}{\bibinfo{person}{Victor Costan}, \bibinfo{person}{Ilia
  Lebedev}, {and} \bibinfo{person}{Srinivas Devadas}.}
  \bibinfo{year}{2016}\natexlab{}.
\newblock \showarticletitle{Sanctum: Minimal hardware extensions for strong
  software isolation}. In \bibinfo{booktitle}{\emph{25th $\{$USENIX$\}$
  Security Symposium ($\{$USENIX$\}$ Security 16)}}. \bibinfo{pages}{857--874}.
\newblock


\bibitem[\protect\citeauthoryear{Crosby and Wallach}{Crosby and
  Wallach}{2009}]%
        {crosby2009efficient}
\bibfield{author}{\bibinfo{person}{Scott~A Crosby} {and} \bibinfo{person}{Dan~S
  Wallach}.} \bibinfo{year}{2009}\natexlab{}.
\newblock \showarticletitle{Efficient Data Structures For Tamper-Evident
  Logging.}. In \bibinfo{booktitle}{\emph{USENIX Security Symposium}}.
  \bibinfo{pages}{317--334}.
\newblock


\bibitem[\protect\citeauthoryear{Das, Eckey, Frassetto, Gens,
  Host{\'a}kov{\'a}, Jauernig, Faust, and Sadeghi}{Das et~al\mbox{.}}{2019}]%
        {das2019fastkitten}
\bibfield{author}{\bibinfo{person}{Poulami Das}, \bibinfo{person}{Lisa Eckey},
  \bibinfo{person}{Tommaso Frassetto}, \bibinfo{person}{David Gens},
  \bibinfo{person}{Kristina Host{\'a}kov{\'a}}, \bibinfo{person}{Patrick
  Jauernig}, \bibinfo{person}{Sebastian Faust}, {and}
  \bibinfo{person}{Ahmad-Reza Sadeghi}.} \bibinfo{year}{2019}\natexlab{}.
\newblock \showarticletitle{$\{$FastKitten$\}$: Practical Smart Contracts on
  Bitcoin}. In \bibinfo{booktitle}{\emph{28th USENIX Security Symposium (USENIX
  Security 19)}}. \bibinfo{pages}{801--818}.
\newblock


\bibitem[\protect\citeauthoryear{Dcunha, Patel, Sawant, Kulkarni, and
  Shirole}{Dcunha et~al\mbox{.}}{2021}]%
        {2021HashTimeLocks}
\bibfield{author}{\bibinfo{person}{Snoviya Dcunha}, \bibinfo{person}{Srushti
  Patel}, \bibinfo{person}{Shravani Sawant}, \bibinfo{person}{Varsha Kulkarni},
  {and} \bibinfo{person}{Mahesh Shirole}.} \bibinfo{year}{2021}\natexlab{}.
\newblock \showarticletitle{Blockchain Interoperability Using Hash Time Locks}.
  In \bibinfo{booktitle}{\emph{Proceeding of Fifth International Conference on
  Microelectronics, Computing and Communication Systems}},
  \bibfield{editor}{\bibinfo{person}{Vijay Nath} {and} \bibinfo{person}{J.~K.
  Mandal}} (Eds.). \bibinfo{publisher}{Springer Singapore},
  \bibinfo{address}{Singapore}, \bibinfo{pages}{475--487}.
\newblock
\showISBNx{978-981-16-0275-7}


\bibitem[\protect\citeauthoryear{{Digital Euro Association}}{{Digital Euro
  Association}}{[n.\,d.]}]%
        {deass_cbdc}
\bibfield{author}{\bibinfo{person}{{Digital Euro Association}}.}
  \bibinfo{year}{[n.\,d.]}\natexlab{}.
\newblock \bibinfo{title}{{Central Bank Digital Currency}}.
\newblock
  \bibinfo{howpublished}{\url{https://home.digital-euro-association.de/cbdc/en}}.
\newblock


\bibitem[\protect\citeauthoryear{{Digital Euro Association}}{{Digital Euro
  Association}}{2022}]%
        {cbdc-manifesto}
\bibfield{author}{\bibinfo{person}{{Digital Euro Association}}.}
  \bibinfo{year}{2022}\natexlab{}.
\newblock \bibinfo{title}{{The CBDC Manifesto}}.
\newblock
\newblock
\urldef\tempurl%
\url{https://cbdcmanifesto.com/}
\showURL{%
\tempurl}


\bibitem[\protect\citeauthoryear{Ehrenfeld}{Ehrenfeld}{2022}]%
        {swift_link}
\bibfield{author}{\bibinfo{person}{Jonathan Ehrenfeld}.}
  \bibinfo{year}{2022}\natexlab{}.
\newblock \bibinfo{title}{SWIFT partners with Chainlink for cross-chain crypto
  transfer project}.
\newblock
\newblock
\urldef\tempurl%
\url{https://blog.chain.link/smartcon-2022-recap/}
\showURL{%
\tempurl}


\bibitem[\protect\citeauthoryear{Enclave}{Enclave}{2019}]%
        {Keystone-enclave}
\bibfield{author}{\bibinfo{person}{Keystone Enclave}.}
  \bibinfo{year}{2019}\natexlab{}.
\newblock \bibinfo{title}{Keystone: An Open Framework for Architecting Trusted
  Execution Environments}.
\newblock \bibinfo{howpublished}{\url{https://keystone-enclave.github.io/}}.
\newblock


\bibitem[\protect\citeauthoryear{Espel, Katz, and Robin}{Espel
  et~al\mbox{.}}{2017}]%
        {espel2017proposal}
\bibfield{author}{\bibinfo{person}{Thomas Espel}, \bibinfo{person}{Laurent
  Katz}, {and} \bibinfo{person}{Guillaume Robin}.}
  \bibinfo{year}{2017}\natexlab{}.
\newblock \showarticletitle{Proposal for Protocol on a Quorum Blockchain with
  Zero Knowledge.}
\newblock   \bibinfo{volume}{2017} (\bibinfo{year}{2017}).
\newblock


\bibitem[\protect\citeauthoryear{Gilad, Hemo, Micali, Vlachos, and
  Zeldovich}{Gilad et~al\mbox{.}}{2017}]%
        {gilad2017algorand}
\bibfield{author}{\bibinfo{person}{Yossi Gilad}, \bibinfo{person}{Rotem Hemo},
  \bibinfo{person}{Silvio Micali}, \bibinfo{person}{Georgios Vlachos}, {and}
  \bibinfo{person}{Nickolai Zeldovich}.} \bibinfo{year}{2017}\natexlab{}.
\newblock \showarticletitle{Algorand: Scaling byzantine agreements for
  cryptocurrencies}. In \bibinfo{booktitle}{\emph{SOSP}}.
\newblock


\bibitem[\protect\citeauthoryear{Gruss, Lettner, Schuster, Ohrimenko, Haller,
  and Costa}{Gruss et~al\mbox{.}}{2017}]%
        {gruss2017strong}
\bibfield{author}{\bibinfo{person}{Daniel Gruss}, \bibinfo{person}{Julian
  Lettner}, \bibinfo{person}{Felix Schuster}, \bibinfo{person}{Olya Ohrimenko},
  \bibinfo{person}{Istvan Haller}, {and} \bibinfo{person}{Manuel Costa}.}
  \bibinfo{year}{2017}\natexlab{}.
\newblock \showarticletitle{Strong and efficient cache side-channel protection
  using hardware transactional memory}. In \bibinfo{booktitle}{\emph{26th
  $\{$USENIX$\}$ Security Symposium ($\{$USENIX$\}$ Security 17)}}.
  \bibinfo{pages}{217--233}.
\newblock


\bibitem[\protect\citeauthoryear{Hoekstra, Lal, Pappachan, Phegade, and
  Del~Cuvillo}{Hoekstra et~al\mbox{.}}{2013}]%
        {hoekstra2013using}
\bibfield{author}{\bibinfo{person}{Matthew Hoekstra}, \bibinfo{person}{Reshma
  Lal}, \bibinfo{person}{Pradeep Pappachan}, \bibinfo{person}{Vinay Phegade},
  {and} \bibinfo{person}{Juan Del~Cuvillo}.} \bibinfo{year}{2013}\natexlab{}.
\newblock \showarticletitle{Using innovative instructions to create trustworthy
  software solutions.}
\newblock \bibinfo{journal}{\emph{HASP@ ISCA}}  \bibinfo{volume}{11}
  (\bibinfo{year}{2013}).
\newblock


\bibitem[\protect\citeauthoryear{Homoliak and Szalachowski}{Homoliak and
  Szalachowski}{2020}]%
        {homoliak2020aquareum}
\bibfield{author}{\bibinfo{person}{Ivan Homoliak} {and} \bibinfo{person}{Pawel
  Szalachowski}.} \bibinfo{year}{2020}\natexlab{}.
\newblock \bibinfo{title}{Aquareum: A Centralized Ledger Enhanced with
  Blockchain and Trusted Computing}.
\newblock
\newblock
\showeprint[arxiv]{2005.13339}~[cs.CR]


\bibitem[\protect\citeauthoryear{{Hyperledger Foundation}}{{Hyperledger
  Foundation}}{2022}]%
        {hyperledger-github}
\bibfield{author}{\bibinfo{person}{{Hyperledger Foundation}}.}
  \bibinfo{year}{2022}\natexlab{}.
\newblock \bibinfo{title}{{Hyperledger}}.
\newblock \bibinfo{howpublished}{\url{https://github.com/hyperledger}}.
\newblock


\bibitem[\protect\citeauthoryear{{Intel}}{{Intel}}{2018}]%
        {intel-sgx-response}
\bibfield{author}{\bibinfo{person}{{Intel}}.} \bibinfo{year}{2018}\natexlab{}.
\newblock \bibinfo{title}{{Resources and Response to Side Channel L1 Terminal
  Fault}}.
\newblock
\newblock
\urldef\tempurl%
\url{https://www.intel.com/content/www/us/en/architecture-and-technology/l1tf.html}
\showURL{%
\tempurl}


\bibitem[\protect\citeauthoryear{Jin and Xia}{Jin and Xia}{2022}]%
        {2022cbdctypes}
\bibfield{author}{\bibinfo{person}{Si~Yuan Jin} {and} \bibinfo{person}{Yong
  Xia}.} \bibinfo{year}{2022}\natexlab{}.
\newblock \showarticletitle{CEV Framework: A Central Bank Digital Currency
  Evaluation and Verification Framework With a Focus on Consensus Algorithms
  and Operating Architectures}.
\newblock \bibinfo{journal}{\emph{IEEE Access}}  \bibinfo{volume}{10}
  (\bibinfo{year}{2022}), \bibinfo{pages}{63698--63714}.
\newblock
\urldef\tempurl%
\url{https://doi.org/10.1109/ACCESS.2022.3183092}
\showDOI{\tempurl}


\bibitem[\protect\citeauthoryear{Kiayias, Russell, David, and
  Oliynykov}{Kiayias et~al\mbox{.}}{2017}]%
        {kiayias2017ouroboros}
\bibfield{author}{\bibinfo{person}{Aggelos Kiayias},
  \bibinfo{person}{Al`exander Russell}, \bibinfo{person}{Bernardo David}, {and}
  \bibinfo{person}{Roman Oliynykov}.} \bibinfo{year}{2017}\natexlab{}.
\newblock \showarticletitle{Ouroboros: A provably secure proof-of-stake
  blockchain protocol}. In \bibinfo{booktitle}{\emph{CRYPTO'17}}.
\newblock


\bibitem[\protect\citeauthoryear{Kiff, Alwazir, Davidovic, Farias, Khan,
  Khiaonarong, Malaika, Monroe, Sugimoto, Tourpe, et~al\mbox{.}}{Kiff
  et~al\mbox{.}}{2020}]%
        {kiff2020survey}
\bibfield{author}{\bibinfo{person}{Mr~John Kiff}, \bibinfo{person}{Jihad
  Alwazir}, \bibinfo{person}{Sonja Davidovic}, \bibinfo{person}{Aquiles
  Farias}, \bibinfo{person}{Mr~Ashraf Khan}, \bibinfo{person}{Mr~Tanai
  Khiaonarong}, \bibinfo{person}{Majid Malaika}, \bibinfo{person}{Mr~Hunter~K
  Monroe}, \bibinfo{person}{Nobu Sugimoto}, \bibinfo{person}{Herv{\'e} Tourpe},
  {et~al\mbox{.}}} \bibinfo{year}{2020}\natexlab{}.
\newblock \showarticletitle{A survey of research on retail central bank digital
  currency}.
\newblock  (\bibinfo{year}{2020}).
\newblock


\bibitem[\protect\citeauthoryear{Kishi}{Kishi}{2019}]%
        {kishi2019project}
\bibfield{author}{\bibinfo{person}{Michinobu Kishi}.}
  \bibinfo{year}{2019}\natexlab{}.
\newblock \bibinfo{title}{Project Stella and the impacts of fintech on
  financial infrastructures in Japan}.
\newblock
\newblock


\bibitem[\protect\citeauthoryear{Koziuk}{Koziuk}{2021}]%
        {Koziuk202112}
\bibfield{author}{\bibinfo{person}{V. Koziuk}.}
  \bibinfo{year}{2021}\natexlab{}.
\newblock \showarticletitle{Confidence in digital money: Are central banks more
  trusted than age is matter?}
\newblock  \bibinfo{volume}{18}, \bibinfo{number}{1} (\bibinfo{year}{2021}),
  \bibinfo{pages}{12--32}.
\newblock


\bibitem[\protect\citeauthoryear{Lacoste and Lefebvre}{Lacoste and
  Lefebvre}{2023}]%
        {lacoste2023trusted}
\bibfield{author}{\bibinfo{person}{Marc Lacoste} {and} \bibinfo{person}{Vincent
  Lefebvre}.} \bibinfo{year}{2023}\natexlab{}.
\newblock \showarticletitle{Trusted Execution Environments for Telecoms:
  Strengths, Weaknesses, Opportunities, and Threats}.
\newblock \bibinfo{journal}{\emph{IEEE Security \& Privacy}}
  (\bibinfo{year}{2023}).
\newblock


\bibitem[\protect\citeauthoryear{Lagarde}{Lagarde}{2022}]%
        {cbdc_privacy_disc}
\bibfield{author}{\bibinfo{person}{Christine Lagarde}.}
  \bibinfo{year}{2022}\natexlab{}.
\newblock \bibinfo{title}{High level conference: Towards a legislative
  framework enabling a digital euro for citizens and businesses}.
\newblock
\newblock
\urldef\tempurl%
\url{https://www.ecb.europa.eu/press/key/date/2022/html/ecb.sp221107~dcc0cd8ed9.en.html}
\showURL{%
\tempurl}


\bibitem[\protect\citeauthoryear{Lan et~al\mbox{.}}{Lan et~al\mbox{.}}{2021}]%
        {lan2021trustcross}
\bibfield{author}{\bibinfo{person}{Ying Lan} {et~al\mbox{.}}}
  \bibinfo{year}{2021}\natexlab{}.
\newblock \showarticletitle{TrustCross: Enabling Confidential Interoperability
  across Blockchains Using Trusted Hardware}. In \bibinfo{booktitle}{\emph{2021
  4th International Conference on Blockchain Technology and Applications}}
  (Xi'an, China) \emph{(\bibinfo{series}{ICBTA 2021})}.
  \bibinfo{publisher}{Association for Computing Machinery},
  \bibinfo{address}{New York, NY, USA}, \bibinfo{pages}{17–23}.
\newblock
\showISBNx{9781450387460}
\urldef\tempurl%
\url{https://doi.org/10.1145/3510487.3510491}
\showDOI{\tempurl}


\bibitem[\protect\citeauthoryear{Lind, Eyal, Kelbert, Naor, Pietzuch, and
  Sirer}{Lind et~al\mbox{.}}{2017}]%
        {lind2017teechain}
\bibfield{author}{\bibinfo{person}{Joshua Lind}, \bibinfo{person}{Ittay Eyal},
  \bibinfo{person}{Florian Kelbert}, \bibinfo{person}{Oded Naor},
  \bibinfo{person}{Peter Pietzuch}, {and} \bibinfo{person}{Emin~G{\"u}n
  Sirer}.} \bibinfo{year}{2017}\natexlab{}.
\newblock \showarticletitle{Teechain: Scalable blockchain payments using
  trusted execution environments}.
\newblock \bibinfo{journal}{\emph{arXiv preprint arXiv:1707.05454}}
  (\bibinfo{year}{2017}).
\newblock


\bibitem[\protect\citeauthoryear{Lipp, Kogler, Oswald, Schwarz, Easdon,
  Canella, and Gruss}{Lipp et~al\mbox{.}}{2021}]%
        {Lipp2021Platypus}
\bibfield{author}{\bibinfo{person}{Moritz Lipp}, \bibinfo{person}{Andreas
  Kogler}, \bibinfo{person}{David Oswald}, \bibinfo{person}{Michael Schwarz},
  \bibinfo{person}{Catherine Easdon}, \bibinfo{person}{Claudio Canella}, {and}
  \bibinfo{person}{Daniel Gruss}.} \bibinfo{year}{2021}\natexlab{}.
\newblock \showarticletitle{{PLATYPUS: Software-based Power Side-Channel
  Attacks on x86}}. In \bibinfo{booktitle}{\emph{2021 IEEE Symposium on
  Security and Privacy (SP)}}. IEEE.
\newblock


\bibitem[\protect\citeauthoryear{Madine, Salah, Jayaraman, Al-Hammadi, Arshad,
  and Yaqoob}{Madine et~al\mbox{.}}{2021}]%
        {2021aapXChain}
\bibfield{author}{\bibinfo{person}{Mohammad Madine}, \bibinfo{person}{Khaled
  Salah}, \bibinfo{person}{Raja Jayaraman}, \bibinfo{person}{Yousof
  Al-Hammadi}, \bibinfo{person}{Junaid Arshad}, {and} \bibinfo{person}{Ibrar
  Yaqoob}.} \bibinfo{year}{2021}\natexlab{}.
\newblock \showarticletitle{appXchain: Application-Level Interoperability for
  Blockchain Networks}.
\newblock \bibinfo{journal}{\emph{IEEE Access}}  \bibinfo{volume}{9}
  (\bibinfo{year}{2021}), \bibinfo{pages}{87777--87791}.
\newblock
\urldef\tempurl%
\url{https://doi.org/10.1109/ACCESS.2021.3089603}
\showDOI{\tempurl}


\bibitem[\protect\citeauthoryear{McKeen, Alexandrovich, Berenzon, Rozas, Shafi,
  Shanbhogue, and Savagaonkar}{McKeen et~al\mbox{.}}{2013}]%
        {mckeen2013innovative}
\bibfield{author}{\bibinfo{person}{Frank McKeen}, \bibinfo{person}{Ilya
  Alexandrovich}, \bibinfo{person}{Alex Berenzon}, \bibinfo{person}{Carlos~V
  Rozas}, \bibinfo{person}{Hisham Shafi}, \bibinfo{person}{Vedvyas Shanbhogue},
  {and} \bibinfo{person}{Uday~R Savagaonkar}.} \bibinfo{year}{2013}\natexlab{}.
\newblock \showarticletitle{Innovative instructions and software model for
  isolated execution.}
\newblock \bibinfo{journal}{\emph{Hasp@isca}} \bibinfo{volume}{10},
  \bibinfo{number}{1} (\bibinfo{year}{2013}).
\newblock


\bibitem[\protect\citeauthoryear{Merkle}{Merkle}{1989}]%
        {merkle1989certified}
\bibfield{author}{\bibinfo{person}{Ralph~C Merkle}.}
  \bibinfo{year}{1989}\natexlab{}.
\newblock \showarticletitle{A certified digital signature}. In
  \bibinfo{booktitle}{\emph{Conference on the Theory and Application of
  Cryptology}}. Springer, \bibinfo{pages}{218--238}.
\newblock


\bibitem[\protect\citeauthoryear{Micali}{Micali}{2003}]%
        {micali2003simple}
\bibfield{author}{\bibinfo{person}{Silvio Micali}.}
  \bibinfo{year}{2003}\natexlab{}.
\newblock \showarticletitle{Simple and fast optimistic protocols for fair
  electronic exchange}. In \bibinfo{booktitle}{\emph{Proceedings of the
  twenty-second annual symposium on Principles of distributed computing}}. ACM,
  \bibinfo{pages}{12--19}.
\newblock


\bibitem[\protect\citeauthoryear{{Microsoft}}{{Microsoft}}{2020}]%
        {eEVM-Microsoft}
\bibfield{author}{\bibinfo{person}{{Microsoft}}.}
  \bibinfo{year}{2020}\natexlab{}.
\newblock \bibinfo{title}{{Enclave EVM}}.
\newblock
\newblock
\urldef\tempurl%
\url{https://github.com/microsoft/eEVM}
\showURL{%
\tempurl}


\bibitem[\protect\citeauthoryear{Miller, Xia, Croman, Shi, and Song}{Miller
  et~al\mbox{.}}{2016}]%
        {miller2016honey}
\bibfield{author}{\bibinfo{person}{Andrew Miller}, \bibinfo{person}{Yu Xia},
  \bibinfo{person}{Kyle Croman}, \bibinfo{person}{Elaine Shi}, {and}
  \bibinfo{person}{Dawn Song}.} \bibinfo{year}{2016}\natexlab{}.
\newblock \showarticletitle{The honeybadger of BFT protocols}. In
  \bibinfo{booktitle}{\emph{ACM CCS}}.
\newblock


\bibitem[\protect\citeauthoryear{Mohanty, Anand, Aljahdali, and Villar}{Mohanty
  et~al\mbox{.}}{2022}]%
        {mohanty2022blockchain}
\bibfield{author}{\bibinfo{person}{Debasis Mohanty}, \bibinfo{person}{Divya
  Anand}, \bibinfo{person}{Hani~Moaiteq Aljahdali}, {and}
  \bibinfo{person}{Santos~Gracia Villar}.} \bibinfo{year}{2022}\natexlab{}.
\newblock \showarticletitle{Blockchain Interoperability: Towards a Sustainable
  Payment System}.
\newblock \bibinfo{journal}{\emph{Sustainability}} \bibinfo{volume}{14},
  \bibinfo{number}{2} (\bibinfo{year}{2022}), \bibinfo{pages}{913}.
\newblock


\bibitem[\protect\citeauthoryear{{Monetary Authority of Singapore}}{{Monetary
  Authority of Singapore}}{2016}]%
        {ubin}
\bibfield{author}{\bibinfo{person}{{Monetary Authority of Singapore}}.}
  \bibinfo{year}{2016}\natexlab{}.
\newblock \bibinfo{title}{{Project Ubin}}.
\newblock
\newblock
\urldef\tempurl%
\url{https://www.mas.gov.sg/schemes-and-initiatives/Project-Ubin}
\showURL{%
\tempurl}


\bibitem[\protect\citeauthoryear{Murdock, Oswald, Garcia, Van~Bulck, Gruss, and
  Piessens}{Murdock et~al\mbox{.}}{2020}]%
        {Murdock2019plundervolt}
\bibfield{author}{\bibinfo{person}{Kit Murdock}, \bibinfo{person}{David
  Oswald}, \bibinfo{person}{Flavio~D. Garcia}, \bibinfo{person}{Jo Van~Bulck},
  \bibinfo{person}{Daniel Gruss}, {and} \bibinfo{person}{Frank Piessens}.}
  \bibinfo{year}{2020}\natexlab{}.
\newblock \showarticletitle{{Plundervolt}: Software-based Fault Injection
  Attacks against Intel SGX}. In \bibinfo{booktitle}{\emph{{Proceedings of the
  41st IEEE Symposium on Security and Privacy (S\&P'20)}}}.
\newblock


\bibitem[\protect\citeauthoryear{Musungate, Candan, Cabuk, and
  Dalkilic}{Musungate et~al\mbox{.}}{2019}]%
        {2019SideChains}
\bibfield{author}{\bibinfo{person}{Blessing Musungate}, \bibinfo{person}{Busra
  Candan}, \bibinfo{person}{Umut Cabuk}, {and} \bibinfo{person}{Gokhan
  Dalkilic}.} \bibinfo{year}{2019}\natexlab{}.
\newblock \showarticletitle{Sidechains: Highlights and Challenges}.
\newblock
\urldef\tempurl%
\url{https://doi.org/10.1109/ASYU48272.2019.8946384}
\showDOI{\tempurl}


\bibitem[\protect\citeauthoryear{Nakamoto}{Nakamoto}{2008}]%
        {nakamoto2008bitcoin}
\bibfield{author}{\bibinfo{person}{Satoshi Nakamoto}.}
  \bibinfo{year}{2008}\natexlab{}.
\newblock \bibinfo{title}{Bitcoin: A peer-to-peer electronic cash system}.
\newblock
\newblock


\bibitem[\protect\citeauthoryear{N{\'a}{\~n}ez~Alonso, Echarte~Fern{\'a}ndez,
  Sanz~Bas, and Kaczmarek}{N{\'a}{\~n}ez~Alonso et~al\mbox{.}}{2020}]%
        {nanez2020reasons}
\bibfield{author}{\bibinfo{person}{Sergio~Luis N{\'a}{\~n}ez~Alonso},
  \bibinfo{person}{Miguel~{\'A}ngel Echarte~Fern{\'a}ndez},
  \bibinfo{person}{David Sanz~Bas}, {and} \bibinfo{person}{Jaros{\l}aw
  Kaczmarek}.} \bibinfo{year}{2020}\natexlab{}.
\newblock \showarticletitle{Reasons fostering or discouraging the
  implementation of central bank-backed digital currency: A review}.
\newblock \bibinfo{journal}{\emph{Economies}} \bibinfo{volume}{8},
  \bibinfo{number}{2} (\bibinfo{year}{2020}), \bibinfo{pages}{41}.
\newblock


\bibitem[\protect\citeauthoryear{Qasse, Abu~Talib, and Nasir}{Qasse
  et~al\mbox{.}}{2019}]%
        {qasse2019inter}
\bibfield{author}{\bibinfo{person}{Ilham~A Qasse}, \bibinfo{person}{Manar
  Abu~Talib}, {and} \bibinfo{person}{Qassim Nasir}.}
  \bibinfo{year}{2019}\natexlab{}.
\newblock \showarticletitle{Inter blockchain communication: A survey}. In
  \bibinfo{booktitle}{\emph{Proceedings of the ArabWIC 6th Annual International
  Conference Research Track}}. \bibinfo{pages}{1--6}.
\newblock


\bibitem[\protect\citeauthoryear{Sebastião, da~Cunha, and Godinho}{Sebastião
  et~al\mbox{.}}{2021}]%
        {Sebastiao2021305}
\bibfield{author}{\bibinfo{person}{H.M.C.V. Sebastião},
  \bibinfo{person}{P.J.O.R. da Cunha}, {and} \bibinfo{person}{P.M.C. Godinho}.}
  \bibinfo{year}{2021}\natexlab{}.
\newblock \showarticletitle{Cryptocurrencies and blockchain. Overview and
  future perspectives}.
\newblock  \bibinfo{volume}{21}, \bibinfo{number}{3} (\bibinfo{year}{2021}),
  \bibinfo{pages}{305--342}.
\newblock


\bibitem[\protect\citeauthoryear{Seo, Lee, Kim, Shih, Shin, Han, and Kim}{Seo
  et~al\mbox{.}}{2017}]%
        {seo2017sgx}
\bibfield{author}{\bibinfo{person}{Jaebaek Seo}, \bibinfo{person}{Byoungyoung
  Lee}, \bibinfo{person}{Seong~Min Kim}, \bibinfo{person}{Ming-Wei Shih},
  \bibinfo{person}{Insik Shin}, \bibinfo{person}{Dongsu Han}, {and}
  \bibinfo{person}{Taesoo Kim}.} \bibinfo{year}{2017}\natexlab{}.
\newblock \showarticletitle{SGX-Shield: Enabling Address Space Layout
  Randomization for SGX Programs.}. In \bibinfo{booktitle}{\emph{NDSS}}.
\newblock


\bibitem[\protect\citeauthoryear{Sethaput and Innet}{Sethaput and
  Innet}{2021}]%
        {sethaput2021blockchain}
\bibfield{author}{\bibinfo{person}{Vijak Sethaput} {and}
  \bibinfo{person}{Supachate Innet}.} \bibinfo{year}{2021}\natexlab{}.
\newblock \showarticletitle{Blockchain Application for Central Bank Digital
  Currencies (CBDC)}. In \bibinfo{booktitle}{\emph{2021 Third International
  Conference on Blockchain Computing and Applications (BCCA)}}. IEEE,
  \bibinfo{pages}{3--10}.
\newblock


\bibitem[\protect\citeauthoryear{Shih, Lee, Kim, and Peinado}{Shih
  et~al\mbox{.}}{2017}]%
        {shih2017t}
\bibfield{author}{\bibinfo{person}{Ming-Wei Shih}, \bibinfo{person}{Sangho
  Lee}, \bibinfo{person}{Taesoo Kim}, {and} \bibinfo{person}{Marcus Peinado}.}
  \bibinfo{year}{2017}\natexlab{}.
\newblock \showarticletitle{T-SGX: Eradicating Controlled-Channel Attacks
  Against Enclave Programs.}. In \bibinfo{booktitle}{\emph{NDSS}}.
\newblock


\bibitem[\protect\citeauthoryear{Siris, Nikander, Voulgaris, Fotiou, Lagutin,
  and Polyzos}{Siris et~al\mbox{.}}{2019}]%
        {2019Interledger}
\bibfield{author}{\bibinfo{person}{Vasilios Siris}, \bibinfo{person}{Pekka
  Nikander}, \bibinfo{person}{Spyros Voulgaris}, \bibinfo{person}{Nikos
  Fotiou}, \bibinfo{person}{Dmitrij Lagutin}, {and} \bibinfo{person}{George
  Polyzos}.} \bibinfo{year}{2019}\natexlab{}.
\newblock \showarticletitle{Interledger Approaches}.
\newblock \bibinfo{journal}{\emph{IEEE Access}}  \bibinfo{volume}{PP}
  (\bibinfo{date}{07} \bibinfo{year}{2019}), \bibinfo{pages}{1--1}.
\newblock
\urldef\tempurl%
\url{https://doi.org/10.1109/ACCESS.2019.2926880}
\showDOI{\tempurl}


\bibitem[\protect\citeauthoryear{{South African Reserve Bank}}{{South African
  Reserve Bank}}{2018}]%
        {Khokha}
\bibfield{author}{\bibinfo{person}{{South African Reserve Bank}}.}
  \bibinfo{year}{2018}\natexlab{}.
\newblock \bibinfo{title}{{Project Khokha: Blockchain Case Study for Central
  Banking in South Africa }}.
\newblock
\newblock
\urldef\tempurl%
\url{https://consensys.net/blockchain-use-cases/finance/project-khokha/}
\showURL{%
\tempurl}


\bibitem[\protect\citeauthoryear{Subramanyan, Sinha, Lebedev, Devadas, and
  Seshia}{Subramanyan et~al\mbox{.}}{2017}]%
        {subramanyan2017formal}
\bibfield{author}{\bibinfo{person}{Pramod Subramanyan}, \bibinfo{person}{Rohit
  Sinha}, \bibinfo{person}{Ilia Lebedev}, \bibinfo{person}{Srinivas Devadas},
  {and} \bibinfo{person}{Sanjit~A Seshia}.} \bibinfo{year}{2017}\natexlab{}.
\newblock \showarticletitle{A formal foundation for secure remote execution of
  enclaves}. In \bibinfo{booktitle}{\emph{Proceedings of the 2017 ACM SIGSAC
  Conference on Computer and Communications Security}}.
  \bibinfo{pages}{2435--2450}.
\newblock


\bibitem[\protect\citeauthoryear{{US Federal Reserve}}{{US Federal
  Reserve}}{[n.\,d.]}]%
        {federalreserve}
\bibfield{author}{\bibinfo{person}{{US Federal Reserve}}.}
  \bibinfo{year}{[n.\,d.]}\natexlab{}.
\newblock \bibinfo{title}{{Central Bank Digital Currency}}.
\newblock
  \bibinfo{howpublished}{\url{https://www.federalreserve.gov/central-bank-digital-currency.htm}}.
\newblock


\bibitem[\protect\citeauthoryear{Van~Bulck, Minkin, Weisse, Genkin, Kasikci,
  Piessens, Silberstein, Wenisch, Yarom, and Strackx}{Van~Bulck
  et~al\mbox{.}}{2018}]%
        {van2018foreshadow}
\bibfield{author}{\bibinfo{person}{Jo Van~Bulck}, \bibinfo{person}{Marina
  Minkin}, \bibinfo{person}{Ofir Weisse}, \bibinfo{person}{Daniel Genkin},
  \bibinfo{person}{Baris Kasikci}, \bibinfo{person}{Frank Piessens},
  \bibinfo{person}{Mark Silberstein}, \bibinfo{person}{Thomas~F Wenisch},
  \bibinfo{person}{Yuval Yarom}, {and} \bibinfo{person}{Raoul Strackx}.}
  \bibinfo{year}{2018}\natexlab{}.
\newblock \showarticletitle{Foreshadow: Extracting the keys to the intel
  $\{$SGX$\}$ kingdom with transient out-of-order execution}. In
  \bibinfo{booktitle}{\emph{27th $\{$USENIX$\}$ Security Symposium
  ($\{$USENIX$\}$ Security 18)}}. \bibinfo{pages}{991--1008}.
\newblock


\bibitem[\protect\citeauthoryear{Wang and Nixon}{Wang and Nixon}{2021}]%
        {2021TrustedRelays}
\bibfield{author}{\bibinfo{person}{Gang Wang} {and} \bibinfo{person}{Mark
  Nixon}.} \bibinfo{year}{2021}\natexlab{}.
\newblock \showarticletitle{InterTrust: Towards an Efficient Blockchain
  Interoperability Architecture with Trusted Services}. In
  \bibinfo{booktitle}{\emph{2021 IEEE International Conference on Blockchain
  (Blockchain)}}. \bibinfo{pages}{150--159}.
\newblock
\urldef\tempurl%
\url{https://doi.org/10.1109/Blockchain53845.2021.00029}
\showDOI{\tempurl}


\bibitem[\protect\citeauthoryear{Wang, Wang, and Chen}{Wang
  et~al\mbox{.}}{2023}]%
        {wang2023exploring}
\bibfield{author}{\bibinfo{person}{Gang Wang}, \bibinfo{person}{Qin Wang},
  {and} \bibinfo{person}{Shiping Chen}.} \bibinfo{year}{2023}\natexlab{}.
\newblock \showarticletitle{Exploring Blockchains Interoperability: A
  Systematic Survey}.
\newblock \bibinfo{journal}{\emph{Comput. Surveys}} (\bibinfo{year}{2023}).
\newblock


\bibitem[\protect\citeauthoryear{Wang, Li, Zhao, and Yu}{Wang
  et~al\mbox{.}}{2020}]%
        {wang2020hybridchain}
\bibfield{author}{\bibinfo{person}{Yong Wang}, \bibinfo{person}{June Li},
  \bibinfo{person}{Siyu Zhao}, {and} \bibinfo{person}{Fajiang Yu}.}
  \bibinfo{year}{2020}\natexlab{}.
\newblock \showarticletitle{Hybridchain: A novel architecture for
  confidentiality-preserving and performant permissioned blockchain using
  trusted execution environment}.
\newblock \bibinfo{journal}{\emph{IEEE Access}}  \bibinfo{volume}{8}
  (\bibinfo{year}{2020}), \bibinfo{pages}{190652--190662}.
\newblock


\bibitem[\protect\citeauthoryear{Zhang and Huang}{Zhang and Huang}{2021}]%
        {zhang2021blockchain}
\bibfield{author}{\bibinfo{person}{Tao Zhang} {and} \bibinfo{person}{Zhigang
  Huang}.} \bibinfo{year}{2021}\natexlab{}.
\newblock \showarticletitle{Blockchain and central bank digital currency}.
\newblock \bibinfo{journal}{\emph{ICT Express}} (\bibinfo{year}{2021}).
\newblock


\end{thebibliography}

\appendix

\clearpage

\begin{algorithm} 
	\scriptsize
	\SetKwProg{func}{function}{}{}
	
	$\triangleright$ \textsc{Declaration of types and functions:}\\
	\hspace{1em} \textbf{Header} \{ $ID$, $txsRoot$, $rcpRoot$, $stRoot$\}; \\
	
	\hspace{1em} $\#(r) \rightarrow v$: denotes the version $v$ of $L$ having  $LRoot$ $=$ $r$,\\		
	
	$\triangleright$ \textsc{Variables of TEE:} \\
	\hspace{1em} $SK_{\mathbb{E}}^{tee}, PK_{\mathbb{E}}^{tee}$: keypair of $\mathbb{E}$ under $\Sigma_{tee}$,\\	
	
	\hspace{1em} $SK_{\mathbb{E}}^{pb}, PK_{\mathbb{E}}^{pb}$: keypair of $\mathbb{E}$ under $\Sigma_{pb}$,\\
	\hspace{1em} $hdr_{last} \leftarrow \perp$: the last header created by $\mathbb{E}$,\\		
	\hspace{1em} $LRoot_{pb} \leftarrow \perp$: the last root of $L$ flushed to PB's $\mathbb{IPSC}$,\\
	\hspace{1em} $LRoot_{cur} \leftarrow \perp$: the root of $L \cup blks_p$ (not flushed to PB),\\	
	\hspace{1em} $ID_{cur} \leftarrow 1$: the current version of $L$ (not flushed to PB),\\
	\hspace{1em} $FH_{cur} \leftarrow []$: the frozen hashes cache of the current $L$'s history tree.\\
	\smallskip
	$\triangleright$ \textsc{Declaration of functions:}
	
	\func{$Init$() \textbf{public}} {
		($SK_{\mathbb{E}}^{pb}$, $PK_{\mathbb{E}}^{pb}$)$ \leftarrow\Sigma_{pb}.Keygen()$;\\ 
		($SK_{\mathbb{E}}^{tee}$, $PK_{\mathbb{E}}^{tee}$)$ \leftarrow\Sigma_{tee}.Keygen()$;\\
		
		\textbf{Output}($PK_{\mathbb{E}}^{tee}, PK_{\mathbb{E}}^{pb}$); \\
	}					
	\smallskip
	
	\func{$Exec$($txs[], \partial st^{old}$) \textbf{public}}{
		
		\textbf{assert} $\partial st^{old}.root = hdr_{last}.stRoot$; \\
		
		$\partial st^{new}, rcps, txs_{er} \leftarrow ~~\_\_processTxs(txs, ~\partial st^{old})$;\\
		
		$\sigma \leftarrow \Sigma_{pb}.sign(SK_{\mathbb{E}}^{pb}, (LRoot_{pb}, LRoot_{cur}))$; \\
		\textbf{Output}($LRoot_{pb}, LRoot_{cur}, \partial st^{new}, hdr_{last}, rcps$, $txs_{er}$, $\sigma$); \\	
		
	}
	\smallskip
	
	\func{$Flush$() \textbf{public}}{									
		$LRoot_{pb} \leftarrow LRoot_{cur}$; \Comment{Shift the version of $L$ synchronized with PB.} \\					
	}
	\smallskip
	
	\func{$\_\_processTxs$($txs[], \partial st^{old}$) \textbf{private}}{
		
		$\partial st^{new}, rcps[], txs_{er} \leftarrow$ runVM($txs$, $\partial st^{old}$); \Comment{Run $\mu$-$txs$ in VM.} \\
		$txs \leftarrow txs \setminus txs_{er}$; \Comment{Filter out parsing errors/wrong signatures.} \\
		$hdr \leftarrow ~~\mathbf{Header}(ID_{cur}, MkRoot(txs), MkRoot(rcps), \partial st^{new}.root))$;\\ 
		$hdr_{last} \leftarrow  hdr$; \\
		$LRoot_{cur} \leftarrow \_\_newLRoot(hdr)$; \\
		$ID_{cur} \leftarrow  ID_{cur} + 1$; \\		
		\textbf{return} $\partial st^{new}$, $rcps$, $txs_{er}$; \\
	}
	\smallskip
	
	\func{$\_\_newLRoot(hdr)$ \textbf{private}}{
		$\_\_udpateFH(h(hdr));$ \\
		\textbf{return} $FH_{cur}.ReduceRoot()$;\\
		\Comment{Since $FH_{cur} = \pi^{inc}_{next}$, inc. proof. for 1 element commitment.} \\
	}
	
	\func{$\_\_updateFH(hdrH)$ \textbf{private}}{
		$FH_{cur}.add(hdrH);$ \\
		$l \leftarrow \lfloor log_2(ID_{cur})\rfloor$;\\
		\For{$i \leftarrow 2;\ i \leq 2^{l};\ i \leftarrow 2i$}{
			\If{$0 = ID_{cur} \bmod i $}{
				$FH_{cur}[\text{-}2] \leftarrow h(FH_{cur}[\text{-}2]\ ||\ FH_{cur}[\text{-}1])$\\
				
				$\mathbf{delete} ~~FH_{cur}[\text{-1}]$; \Comment{Remove the last element.}
			}
		}
	}
	
	\caption{The program $prog^{\mathbb{E}}$ of enclave $\mathbb{E}$}
	\label{alg:enclave-VM}
\end{algorithm}

\begin{algorithm} 
	\scriptsize 
	\caption{The program $prog^{\mathbb{IPSC}}$ of $\mathbb{IPSC}$.}\label{alg:IPSC-smart-contract}
	
	\SetKwProg{func}{function}{}{}

	\smallskip
	$\triangleright$ \textsc{Declaration of types and constants:}\\		
	\hspace{1em} \textbf{CensInfo} \{ $\mu$-$etx, \mu$-$equery, status, edata$ \},  \\
	\hspace{1em} $msg$: a current transaction that called $\mathbb{IPSC}$,  \\
	
	\smallskip
	$\triangleright$ \textsc{Declaration of functions:}
	
	\func{$Init$($PK_{\mathbb{E}}^{pb}, PK_{\mathbb{E}}^{tee}, PK_{\mathbb{O}}, ~\diff{\_i_r},~ \diff{[ia \leftarrow \textbf{T}]} $) \textbf{public} }{
		$PK_{\mathbb{E}}^{tee}[].add(PK_{\mathbb{E}}^{tee})$; \Comment{PK of enclave $\mathbb{E}$ under $\Sigma_{tee}$.} \\ 
		$PK_{\mathbb{E}}^{pb}[].add(PK_{\mathbb{E}}^{pb})$; \Comment{PK of enclave $\mathbb{E}$ under $\Sigma_{pb}$.} \\
		$PK_{\mathbb{O}}^{pb} \leftarrow PK_{\mathbb{O}}$; \Comment{PK of operator $\mathbb{O}$ under $\Sigma_{pb}$.} \\
		$LRoot_{pb} \leftarrow \perp$; \Comment{The most recent root hash of $L$ synchronized with $\mathbb{IPSC}$.} \\ 
		$censReqs \leftarrow []$; \Comment{Request that $\mathbb{C}$s wants to resolve publicly.} \\
		\diff{$t_s \leftarrow 0$}; \Comment{The total supply of the instance.}\\
		\diff{$t_i \leftarrow 0$}; \Comment{The total issued tokens by the instance.}\\
		\diff{\textbf{const} $issueAuthority \leftarrow ia$}; \Comment{Token issuance  capability of the instance.} \\ 
		\diff{\textbf{const} $i_r \leftarrow \_i_r$}; \Comment{Max. yearly inflation of the instance.} \\ 
		\diff{\textbf{const} $createdAt \leftarrow timestamp()$}; \Comment{The timestamp of creation a CBDC instance.} \\ 
	}
	
	\func{$snapshotLedger$($root_A, root_B, \diff{\_t_i, \_t_s,}~ \sigma$) \textbf{public} }{
		\Comment{Verify whether msg was signed by $\mathbb{E}$. \hfill} \\
		\textbf{assert} $\Sigma_{pb}.verify((\sigma, PK_{\mathbb{E}}^{pb}[\text{-}1]), (root_A, root_B, \diff{\_t_i, \_t_s}))$;  \\			
		
		\Comment{Snapshot issued tokens and total supply. \hfill} \\
		\diff{
			\If{$issueAuthority$}{
				\textbf{assert} $\_\_meetsInflationRate(\_t_i)$;  \Comment{The code is trivial, and we omit it.} \\
				$t_i \leftarrow \_t_i$;			\\
			}
			\Else{
				\diff{\textbf{assert} $t_i = \_t_i$}; \\ 
			}
		}
		
		\Comment{Verify whether a version transition extends the last one. \hfill} \\
		\If{$LRoot_{pb} = root_A$}{
			$LRoot_{pb} \leftarrow root_{B}$; \Comment{Do a version transition of $L$.} \\
		}	
		
	}
	\smallskip

	\func{$SubmitCensTx$($\mu$-$etx, \sigma_{msg}$) \textbf{public} }{
		\Comment{Called by $\mathbb{C}$ in the case her $\mu$-tx is censored. $\mathbb{C}$ encrypts it by $PK^{tee}_{\mathbb{E}}$. \hfill \hfill}\\		
		accessControl($\sigma_{msg}, msg.PK_{\mathbb{C}}^{pb}$); \\

		$censReqs$.add(\textbf{CensInfo}($\mu$-$etx, \perp, \perp, \perp$)); \\			
		
	}
	\smallskip
	
	\func{$ResolveCensTx(idx_{req}, status, \sigma$) \textbf{public} }{
		\Comment{Called by $\mathbb{O}$ to prove that $\mathbb{C}$'s $\mu$-tx was processed.\hfill \hfill \hfill}\\
		
		\textbf{assert} $idx_{req} < |censReqs|$;\\
		$r \leftarrow censReqs[idx_{req}]$; \\
		
		\textbf{assert} $\Sigma_{pb}.verify((\sigma, PK_{\mathbb{E}}^{pb}[\text{-}1]), ~(h(r.\mu$-$etx), status))$; 	\\ 
		$r.status \leftarrow status$;\\
	}	
	
	\func{$SubmitCensQry$($\mu$-$equery, \sigma_{msg}$) \textbf{public} }{
		\Comment{Called by $\mathbb{C}$ in the case its read query is censored. $\mathbb{C}$ encrypts it by $PK^{tee}_{\mathbb{E}}$.}\\		
		accessControl($msg$, $\sigma_{msg}, msg.PK_{\mathbb{C}}^{pb}$); \\

		$censReqs$.add(\textbf{CensInfo}($\perp, \mu$-$equery, \perp, \perp$)); \\		
		
	}
	\smallskip
	
	\func{$ResolveCensQry(idx_{req}, status, edata, \sigma$) \textbf{public} }{
		\Comment{Called by $\mathbb{O}$ as a response to the $\mathbb{C}$'s censored  read query.\hfill \hfill \hfill}\\		
		\textbf{assert} $idx_{req} < |censReqs|$;\\
		$r \leftarrow censReqs[idx_{req}]$; \\
		
		\textbf{assert} $\Sigma_{pb}.verify((\sigma, PK_{\mathbb{E}}^{pb}[\text{-}1]), (h(r.\mu$-$equery), status, h(edata)))$; 	\\ 
		$r.\{edata \leftarrow edata, status \leftarrow status\}$;\\		
	}
	
	\smallskip
	
	\func{$ReplaceEnc$($PKN_{\mathbb{E}}^{pb}, PKN_{\mathbb{E}}^{tee}, r_{A}, r_{B}, ~\diff{\_t_i, \_t_s,}~  \sigma, \sigma_{msg}$) \textbf{public} }{
		\Comment{Called	 by $\mathbb{O}$ in the case of enclave failure.\hfill}\\
		
		\textbf{assert} $\Sigma_{pb}.verify((\sigma_{msg}, PK_{\mathbb{O}}^{pb}), msg)$;  \Comment{Avoiding MiTM attack.} \\
		$snapshotLedger(r_{A}, r_{B}, ~\diff{\_t_i, \_t_s},~ \sigma)$ ; \Comment{Do a version transition.} \\

		$PK_{\mathbb{E}}^{tee}.add(PKN_{\mathbb{E}}^{tee})$; \Comment{Upon change, $\mathbb{C}s$ make remote attestation.} \\ 
		$PK_{\mathbb{E}}^{pb}.add(PKN_{\mathbb{E}}^{pb})$; \\	
	}

\end{algorithm}

\end{document}